\documentclass[%
 prb,
 amsmath,amssymb,
 reprint,%
]{revtex4-2}

\usepackage{graphicx}
\usepackage{amsmath}
\usepackage{float}
\usepackage{amssymb}
\usepackage{placeins}
\usepackage{array}
\usepackage{color}
\usepackage{physics}
\usepackage{makecell}
\usepackage{siunitx}
\usepackage{xstring}
\usepackage{multirow}
\usepackage{mathtools}
\usepackage{soul} 
\usepackage{changes}
\usepackage{dcolumn}
\usepackage{bm}
\usepackage{subcaption} 

\DeclareUnicodeCharacter{0301}{\'{e}}

\newcommand{\rn}[1]{%
  \textup{\uppercase\expandafter{\romannumeral#1}}%
}

\def\ctcn{\text{C}_\text{2}\text{C}_\text{N}}
\def\ctcb{\text{C}_\text{2}\text{C}_\text{B}}
\def\cbvn{\text{C}_\text{B}\text{V}_\text{N}}

\def\gs{\prescript{2}{0}{A}_2}
\def\esb{\prescript{2}{1}{B}_1} 
\def\est{\prescript{2}{2}{A}_2} 

\def\gw{\text{G}_0\text{W}_0}
\def\gwbse{\text{G}_0\text{W}_0/\text{BSE}}
\definecolor{tcolor}{rgb}{0, 0.5, 0.2}

\begin{document}
\preprint{AIP/123-QED}

\title{$\ctcn$ as a 2 eV Single-Photon Emitter Candidate in Hexagonal Boron Nitride}

\author{Kejun Li}
\affiliation{Department of Physics, University of California, Santa Cruz, CA, 95064, USA}
\author{Tyler J. Smart}
\affiliation{Department of Physics, University of California, Santa Cruz, CA, 95064, USA}
\affiliation{Lawrence Livermore National Laboratory, Livermore, California 94551, USA}
\author{Yuan Ping}
\email{yuanping@ucsc.edu}
\affiliation{Department of Chemistry and Biochemistry, University of California, Santa Cruz, CA, 95064, USA}

\date{\today}

\begin{abstract}
    The generation of single-photon emitters in hexagonal boron nitride around 2 eV emission is experimentally well-recognized; however the atomic nature of these emitters is unknown.
    In this paper, we use first-principles calculations to demonstrate that $\ctcn$ is a possible source of 2 eV single-photon emitter. We showcase the calculations of a complete set of static and dynamical properties related to defects, including exciton-defect couplings and electron-phonon interactions. In particular, we show it is critical to consider nonradiative processes when comparing with experimental lifetime for known 2 eV emitters.
    We find that $\ctcn$ has 
    several key physical properties matching the ones of experimentally observed single-photon emitters.
    These include the zero-phonon line (2.13 eV), Huang-Rhys factor (1.35), photoluminescence lifetime (2.19 ns), phonon-sideband energy (180 meV), and photoluminescence spectrum.
    The identification of defect candidates for 2 eV emission paves the way for controllable single-photon emission generation.    
    %
\end{abstract}

\maketitle

\section{Introduction}
A single-photon emitter is a crucial component in building quantum information technologies, such as linear-optic quantum information processing~\cite{wang2020integrated}, quantum simulation~\cite{wang2019boson} and quantum communication~\cite{barrett2005no}. Hexagonal boron nitride (hBN) is a two-dimensional (2D) material with a wide bandgap ($\sim$6 eV)~\cite{cassabois2016hexagonal, elias2019direct}, and can host stable and bright color centers which possess single-photon emission~\cite{fischer2021controlled}. Meanwhile, the manipulation of the optoelectronic properties of singe-photon emitters has garnered special interest~\cite{zpl_raman_pse_tau_spe_grosso2017tunable, qyli2019near}. These properties imply a great potential of hBN in developing quantum applications. Being able to select and purify single-photon emitters is critical for generating 
controllable and narrow line width single-photon emission. Hence, the identification of the atomic origin of the single-photon emitters is crucial to the development of this field.

Since 2016~\cite{zpl_raman_dw_tau_qy_spe_linpol_tran2016quantummono}, numerous experimental results have been reported on the photoluminescent properties of the single-photon emitters at $\sim$2 eV in hBN~\cite{zpl_raman_hr_tau_spe_tran2016quantum, zpl_pse_tau_spe_martinez2016efficient, zpl_pse_tau_spe_dietrich2018observation, zpl_raman_dw_tau_qy_spe_linpol_tran2016quantummono, zpl_raman_pse_spe_linpol_spinpol_mendelson2020identifying, zpl_pse_spe_hayee2020revealing, zpl_raman_pse_tau_spe_grosso2017tunable, zpl_pse_dw_tau_qy_spe_linpol_nguyen2018nanoassembly, zpl_raman_tau_spe_linpol_xu2018single, zpl_pse_tau_spe_linpol_khatri2020optical, zpl_tau_spe_linpol_lazic2019dynamically, zpl_raman_pse_spe_linpol_liu2020ultrastable, zpl_pse_tau_spe_vogl2018fabrication, zpl_pse_tau_spe_tran2018resonant, zpl_raman_pse_spe_mendelson2019engineering, zpl_pse_hr_tau_spe_linpol_exarhos2017optical, zpl_pse_tau_spe_tran2016robust, zpl_pse_spe_ngoc2018effects, zpl_pse_spe_choi2016engineering, zpl_tau_spe_grosso2020low, zpl_tau_spe_linpol_lazic2019dynamically}.
In TABLE~\ref{tab:exp}, we summarize the range and averaged values of  photoluminescent properties observed in  experiments, including the zero-phonon line (ZPL), phonon-sideband energy (PSE), Huang-Rhys factor (HR factor), and the photoluminescence lifetime ($\tau^{PL}$). Typically, a ${\sim}2$ eV single-photon emitter in hBN exhibits a photoluminescence (PL) spectrum comprised of a sharp ZPL and one or two moderate phonon-sidebands (PSB) at room temperature. The PSE is the energy separation between the ZPL and the first PSB peak. The electron-phonon coupling is estimated by either HR factor ($S$) or Debye-Waller (DW factor$=e^{-S}$)~\cite{neu2011single}. The HR factor is typically $\sim$1, which indicates weak electron-phonon coupling. Furthermore, the PL lifetime is on the order of a few ns, and quantum yield (QY, $\eta$) is reported to be $6{\sim}12$\%~\cite{qyli2019near}. PL lifetime ($\tau^{PL}$) reflects the lifetime of an excited state, determined by radiative ($\tau^{R}$) and nonradiative recombination together ($\tau^{NR}$, $\tau^{PL}=1/(1/\tau^{R}+1/\tau^{NR})$). On the other hand, QY ($\eta$) reflects the proportion of radiative recombination rates with respect to the total recombination rates, and is related to PL lifetime through $\tau^{PL}=\tau^{R}*\eta$. PL lifetime and QY underscore the importance of studying both radiative and nonradiative recombination lifetimes. 
\begin{table}
    \centering
    \caption{Summary of the zero-phonon line (ZPL), phonon-sideband (PSE), Huang-Rhys (HR) factor and photoluminescence (PL) lifetime ($\tau^{\mathrm{PL}}$) for ${\sim}2$ eV single-photon emitters in $h$-BN from experimental measurements and our calculated two carbon defects.}
    \begin{ruledtabular}
    \begin{tabular}{ccccc}
               &  ZPL (eV) 
               &  PSE (meV)
               &  HR factor
               &  $\tau^{\mathrm{PL}}$ (ns) \\
        \hline
        Range(exp.)  &  1.56-2.24\footnotemark[1]
               &  30-200\footnotemark[2]
               &  0.63-1.93\footnotemark[3]
               &  0.38-19.7\footnotemark[4] \\
        Mean(exp.)   &  $2.00\pm0.19$\footnotemark[1]
               &  $156\pm33$\footnotemark[2]  
               &  $1.19\pm0.43$\footnotemark[3]  
               &  $2.6\pm1.5$\footnotemark[4] \\
        $\ctcn$(cal.)&  2.13\footnotemark[5]
               &  180
               &  1.35
               &  2.19\\
        $\ctcb$(cal.)&  1.42\footnotemark[5]
               &  175
               &  1.25
               &  $3.83*10^{2}$\\
    \end{tabular}
    \end{ruledtabular}
    \label{tab:exp}
    \footnotetext[1]{From Ref.~\citenum{zpl_raman_hr_tau_spe_tran2016quantum, zpl_pse_tau_spe_martinez2016efficient, zpl_pse_tau_spe_dietrich2018observation, zpl_pse_linpol_konthasinghe2019rabi, zpl_raman_dw_tau_qy_spe_linpol_tran2016quantummono, zpl_raman_pse_spe_linpol_spinpol_mendelson2020identifying, zpl_pse_spe_hayee2020revealing, zpl_raman_pse_tau_spe_grosso2017tunable, zpl_pse_dw_tau_qy_spe_linpol_nguyen2018nanoassembly, zpl_raman_tau_spe_linpol_xu2018single, zpl_pse_tau_spe_linpol_khatri2020optical, zpl_raman_pse_hr_dw_wang2018photoluminescence, zpl_tau_spe_linpol_lazic2019dynamically, zpl_raman_pse_spe_linpol_liu2020ultrastable, zpl_pse_tau_spe_vogl2018fabrication, zpl_pse_tau_spe_tran2018resonant, zpl_tau_linpol_schell2017coupling, zpl_raman_pse_spe_mendelson2019engineering, zpl_pse_hr_tau_spe_linpol_exarhos2017optical, zpl_pse_tau_spe_tran2016robust, zpl_pse_spe_ngoc2018effects, zpl_pse_spe_choi2016engineering, zpl_tau_spe_grosso2020low, zpl_tau_spe_linpol_lazic2019dynamically}.}
    \footnotetext[2]{From Ref.~\citenum{zpl_pse_tau_spe_martinez2016efficient, zpl_pse_tau_spe_dietrich2018observation, zpl_pse_linpol_konthasinghe2019rabi, zpl_raman_pse_spe_linpol_spinpol_mendelson2020identifying, zpl_pse_spe_hayee2020revealing, zpl_raman_pse_tau_spe_grosso2017tunable, zpl_pse_dw_tau_qy_spe_linpol_nguyen2018nanoassembly, zpl_pse_tau_spe_linpol_khatri2020optical, zpl_raman_pse_hr_dw_wang2018photoluminescence, zpl_raman_pse_spe_linpol_liu2020ultrastable, zpl_pse_tau_spe_vogl2018fabrication, zpl_pse_tau_spe_tran2018resonant, zpl_raman_pse_spe_mendelson2019engineering, zpl_pse_hr_tau_spe_linpol_exarhos2017optical, zpl_pse_tau_spe_tran2016robust, zpl_pse_spe_ngoc2018effects, zpl_pse_spe_choi2016engineering}.}
    \footnotetext[3]{From Ref.~\citenum{zpl_raman_hr_tau_spe_tran2016quantum, zpl_raman_pse_hr_dw_wang2018photoluminescence, zpl_pse_hr_tau_spe_linpol_exarhos2017optical}.}
    \footnotetext[4]{From Ref.~\citenum{zpl_raman_hr_tau_spe_tran2016quantum, zpl_pse_tau_spe_martinez2016efficient, zpl_pse_tau_spe_dietrich2018observation, zpl_raman_dw_tau_qy_spe_linpol_tran2016quantummono, zpl_raman_pse_tau_spe_grosso2017tunable, zpl_pse_dw_tau_qy_spe_linpol_nguyen2018nanoassembly, zpl_raman_tau_spe_linpol_xu2018single, zpl_pse_tau_spe_linpol_khatri2020optical, zpl_pse_tau_spe_vogl2018fabrication, zpl_pse_tau_spe_tran2018resonant, zpl_tau_linpol_schell2017coupling, zpl_pse_hr_tau_spe_linpol_exarhos2017optical, zpl_pse_tau_spe_tran2016robust}. As 19.7 ns in Ref.~\citenum{zpl_tau_linpol_schell2017coupling} is far from the other data points, it is removed when evaluating the mean.}
    \footnotetext[5]{From the excitation energy at $\gwbse$@PBE and taking into account the Franck-Condon shift.}
\end{table}
Finally, many of the reported single-photon emitters ${\sim2}$ eV possess linearly polarized excitation and emission~\cite{zpl_pse_linpol_konthasinghe2019rabi, zpl_raman_dw_tau_qy_spe_linpol_tran2016quantummono, zpl_raman_pse_spe_linpol_spinpol_mendelson2020identifying, zpl_pse_dw_tau_qy_spe_linpol_nguyen2018nanoassembly, zpl_raman_tau_spe_linpol_xu2018single, zpl_pse_tau_spe_linpol_khatri2020optical, zpl_tau_spe_linpol_lazic2019dynamically, zpl_raman_pse_spe_linpol_liu2020ultrastable, zpl_tau_linpol_schell2017coupling, zpl_spe_linpol_spinpol_exarhos2019magnetic, zpl_pse_hr_tau_spe_linpol_exarhos2017optical}, indicating the anisotropic structural symmetry of the corresponding defects~\cite{wu2019dimensionality, yim2020polarization}, i.e. possibly belonging to
$C_{2v}$, $C_2$ or $C_s$ groups.

In terms of the atomic origin of these single-photon emitters, annular dark-field images have shown that carbon substitutions are abundant in hBN~\cite{krivanek2010atom}.
Recently, Mendelson et al.~\cite{zpl_raman_pse_spe_linpol_spinpol_mendelson2020identifying} has identified that various techniques of incorporating carbon into hBN yield $\sim$2 eV single-photon emitters.
Moreover, X-ray photoelectron spectroscopy measurements show that more C-B bonds exist than that of C-N bonds~\cite{zpl_raman_pse_spe_linpol_spinpol_mendelson2020identifying}. This is an evidence that carbon substitution of nitrogen is more likely than carbon substitution of boron.

Theoretically, several defects have been proposed to be possibly responsible for single-photon emission, such as $\cbvn$~\cite{sajid2020vncb,fischer2021controlled}, boron dangling bonds~\cite{turiansky2021impact}, $\mathrm{N_BV_N}$~\cite{wu2019carrier}, and carbon trimers~\cite{jara2021first}.
Based on experimental observations~\cite{zpl_raman_pse_spe_linpol_spinpol_mendelson2020identifying}, we focus on carbon defects in this work. Among the carbon defects, $\cbvn$ has high formation energy, $3{\sim}6$ eV higher than other carbon defects such as carbon dimer and trimers~\cite{cheng2017paramagnetic}. Carbon trimers including $\ctcn$ and $\ctcb$, theoretically proposed by Jara et al.~\cite{jara2021first},  were found energetically favorable and in good agreement with experimental PSE and PL~\cite{jara2021first}. However, some important information is still missing to confirm carbon trimers as a SPE candidate in  Ref.~\citenum{jara2021first}. For example, only the ZPL for the lowest transition was calculated, about 0.4 eV smaller than the mean of ZPL in TABLE~\ref{tab:exp}. And other important properties such as HR factor, PL lifetime and QY, as the experimental characteristics of ${\sim}2$ eV single-photon emitters, were not calculated. Therefore, further study on these properties is desired to unveil the role of carbon trimers as experimentally observed SPEs.

In this paper, electronic and optical  properties of the carbon trimers $\ctcn$ and $\ctcb$ are calculated from both first-principles many-body perturbation theory and DFT levels. We evaluate both static and dynamical properties of radiative and phonon-assisted nonradiative recombination, including exciton-defect and electron-phonon interactions. 
Our results demonstrate that $\ctcn$ has remarkable agreement with experimental observations of 2 eV single photon emitters, including ZPL, PL lifetime, PSE, HR factor, and PL spectrum. We emphasize the importance of comparing all key signatures between theory and experiments for defect identification and validations.   

\section{Computational Methods}
We employ open source plane-wave code Quantum ESPRESSO~\cite{QE} for structural relaxation and phonon calculations of carbon defects in monolayer hBN. We use the optimized norm-conserving Vanderbilt (ONCV) pseudopotentials~\cite{ONCV1} and a 55 Ry wavefunction energy cutoff. We choose a supercell size of $6\times6$ which shows good convergence as tested in Ref.~\citenum{wu2019carrier, Tyler2021-err, Smart2021wk}.  
Charged defect correction is included to eliminate the spurious electrostatic interaction by using the techniques developed in Refs.~\citenum{PingJCP, wu2017first} and implemented in the JDFTx code~\cite{JDFTx}. Total energy, defect formation energy and geometry are obtained with the Perdew-Burke-Ernzerhof (PBE) exchange correlation functional~\cite{PBE1997}. The charged defect formation energy $E_f^q(d)$ with the charge state $q$ is calculated by
\begin{align}
    E_f^q(d) = E_{\mathrm{tot}}^q(d) - E_{\mathrm{tot}}(p) - \sum_iN_i\mu_i + qE_{\mathrm{Fermi}} + E_{\mathrm{corr}},
    \label{eq:formation_energy}
\end{align}
where $E_{\mathrm{tot}}^q(d)$ is the total energy of the charged defect, $E_{\mathrm{tot}}(p)$ is the total energy of the pristine system, $N_i$ is the number of atoms of atomic species $i$ that is added ($N_i>0$) or removed ($N_i<0$), $\mu_i$ is the chemical potential of the atomic species $i$, $E_{\mathrm{Fermi}}$ is the electron chemical potential, and $E_{\mathrm{corr}}$ is the charged defect correction. The chemical potentials of B, N and C are obtained as follows. In the N-rich condition, $\mathrm{\mu^{N-rich}_N}=1/2E_{\mathrm{tot}}\mathrm{(N_2)}$ where $E_{\mathrm{tot}}\mathrm{(N_2)}$ is the total energy of $\mathrm{N_2}$ molecule. In the N-poor condition, $\mathrm{\mu^{N-poor}_B}=E_{\mathrm{tot}}\mathrm{(B)}$ where $E_{\mathrm{tot}}\mathrm{(B)}$ is one atom's total energy in a boron crystal. $\mathrm{\mu^{N-poor}_N}$ and $\mathrm{\mu^{N-rich}_B}$ are calculated according to the constraint $\mu_\mathrm{N}+\mu_\mathrm{B}=\mu_\mathrm{BN}$, where $\mu_\mathrm{BN}$ is the total energy of BN with one unit cell. $\mu_\mathrm{C}$, on the other hand, is one atom's total energy in graphene.

We evaluate electronic structures under GW approximation with the PBE eigenvalues and wavefunctions as the starting point by using the Yambo code~\cite{YAMBO}. The starting wavefunctions at PBE are sufficient for accurate descriptions of current $sp$ defect wavefunctions by comparing with the ones at hybrid functionals (see SI FIG. S1 for wavefunction comparison~\cite{fuchs2007quasiparticle,bechstedt2016many,Smart2021wk}). 
In single-shot GW ($\gw$) calculations in this work, the Godby-Needs plasmon-pole approximation (PPA)~\cite{godby1989metal, oschlies1995gw} is used to calculate the dielectric matrices with the plasmon frequency $\omega_p=27.2$ eV. Then  Bethe-Salpeter Equation (BSE) is further solved on top of the GW approximation to include the electron-hole interaction in absorption spectra. Coulomb truncation for 2D systems~\cite{rozzi2006exact} is applied to the out-of-plane direction, and k-point sampling is set to $3\times3\times1$ for $6\times6\times1$ supercells in all of the $\gw$ and BSE calculations. More details on convergence tests can be found in SI Section \rn2. 

The method in Ref.~\citenum{wu2019dimensionality} based on Fermi's golden rule and solving BSE is applied for calculating radiative lifetime of intra-defect transitions that include exciton-defect coupling. Nonradiative recombination lifetime is also calculated with electron-phonon matrix elements in the static coupling approximation. The details of the nonradiative recombination can be found in Ref.~\citenum{nonrad_alkauskas2014first, wu2019carrier}. The generating function approach, which is detailed introduced in Ref.~\citenum{pl_alkauskas2014first}, is applied to the calculation of PL lineshape. More benchmark and convergence tests on PL calculations can be found in SI Section \rn5.

\section{Results and Discussions}
Our results and discussions are organized as follows. First, we discuss the defect stability and their stable charge states through defect formation energies and charge transition levels. Then, we present the electronic structure, optical spectra, and radiative lifetime of the defects. Next, we show ZPL and nonradiative recombination processes of the intra-defect transitions (transition between two states localized on one isolated defect). We then discuss competition between radiative and nonradiative recombination processes and the resulting PL lifetime. Moreover, the PL lineshape and spectral function with the dominant phonon mode are discussed. Finally, we show a full picture comprised of ZPL, PL lifetimes, PSE, HR factor, and PL spectrum of $\ctcn$ in comparison with experiments.

\subsection{Charged Defect Formation Energy}
When nitrogen and boron atoms are unevenly substituted by atomic impurities, the formation energy of the defects depends on the elemental chemical potentials, i.e. different at the N-rich (B-poor) and N-poor (B-rich) conditions. In the N-rich condition, $\ctcb$ is more likely to form than $\ctcn$ because of the smaller formation energy of $\ctcb$, as shown in FIG.~\ref{eq:formation_energy}(b). On the other hand, $\ctcn$ is more likely to form in the N-poor condition. It is also found that both defects can be stable in charge states of $q=0,\pm1$ at a range of electron chemical potentials within the electronic band gap. Without losing generality, we investigate both $\ctcn$ and $\ctcb$ in all possible charge states of $q=0,\pm1$, and we find that the $q=\pm1$ charge states have ZPL either too large or too small, away from the experimental range (details found in SI TABLE S1). Thus, we focus on the neutral state of the defects. In particular, we find that the neutral $\ctcn$ have good agreement with experiments on all properties of $\sim2$ eV single-photon emitter, so we mainly discuss the neutral $\ctcn$ in the following. The results of $\ctcb$ can be found in SI Section \rn3 and summarized in Table~\ref{tab:exp}. We find the major discrepancy of $\ctcb$ to experiments is in ZPL and PL lifetime. 

\begin{figure}
    \centering
    \includegraphics[width=0.48\textwidth]{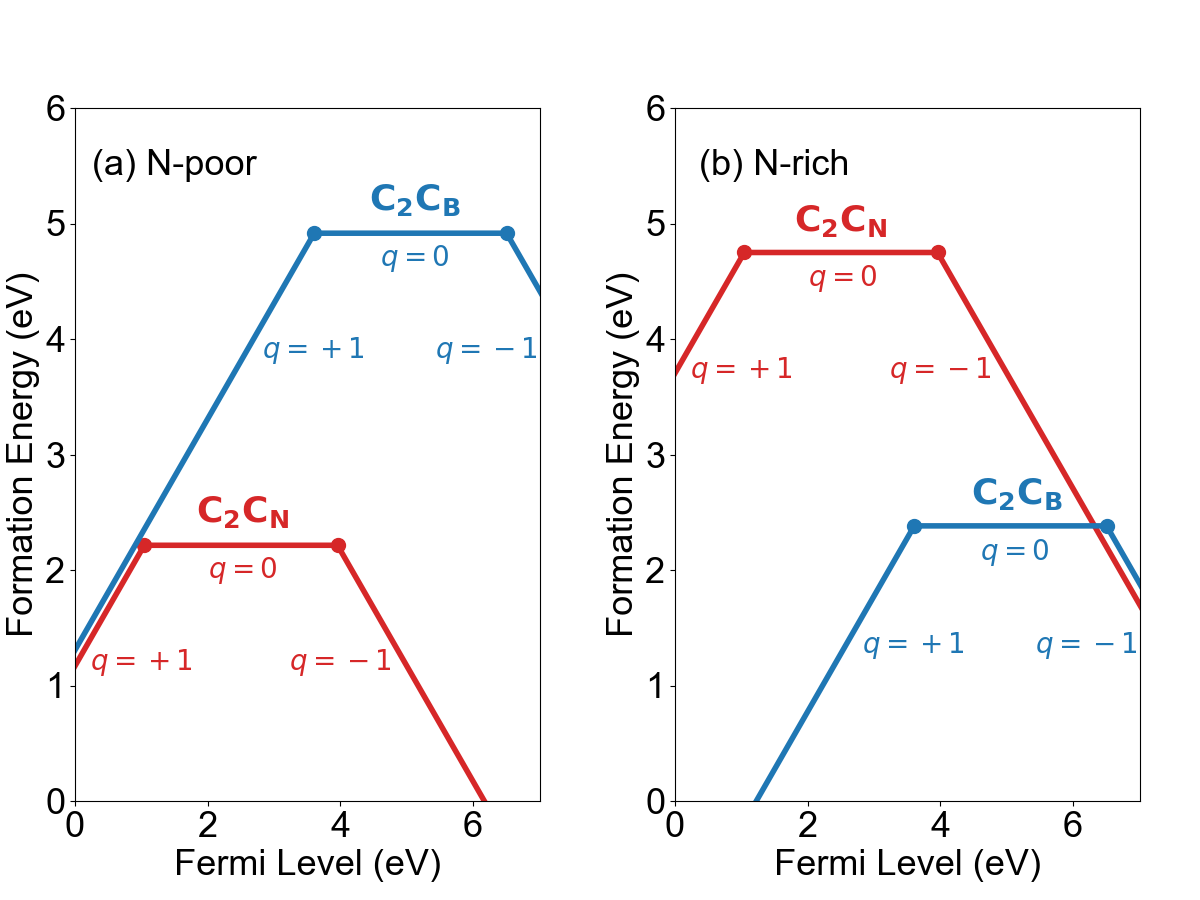}
    \caption{Charged defect formation energy of defects $\ctcn$ and $\ctcb$ as a function of Fermi level at (a) N-poor and (b) N-rich conditions.}
    \label{fig:ctl}
\end{figure}

\subsection{Electronic Structure and Optical Properties}
\begin{figure*}
    \includegraphics[width=0.9\textwidth]{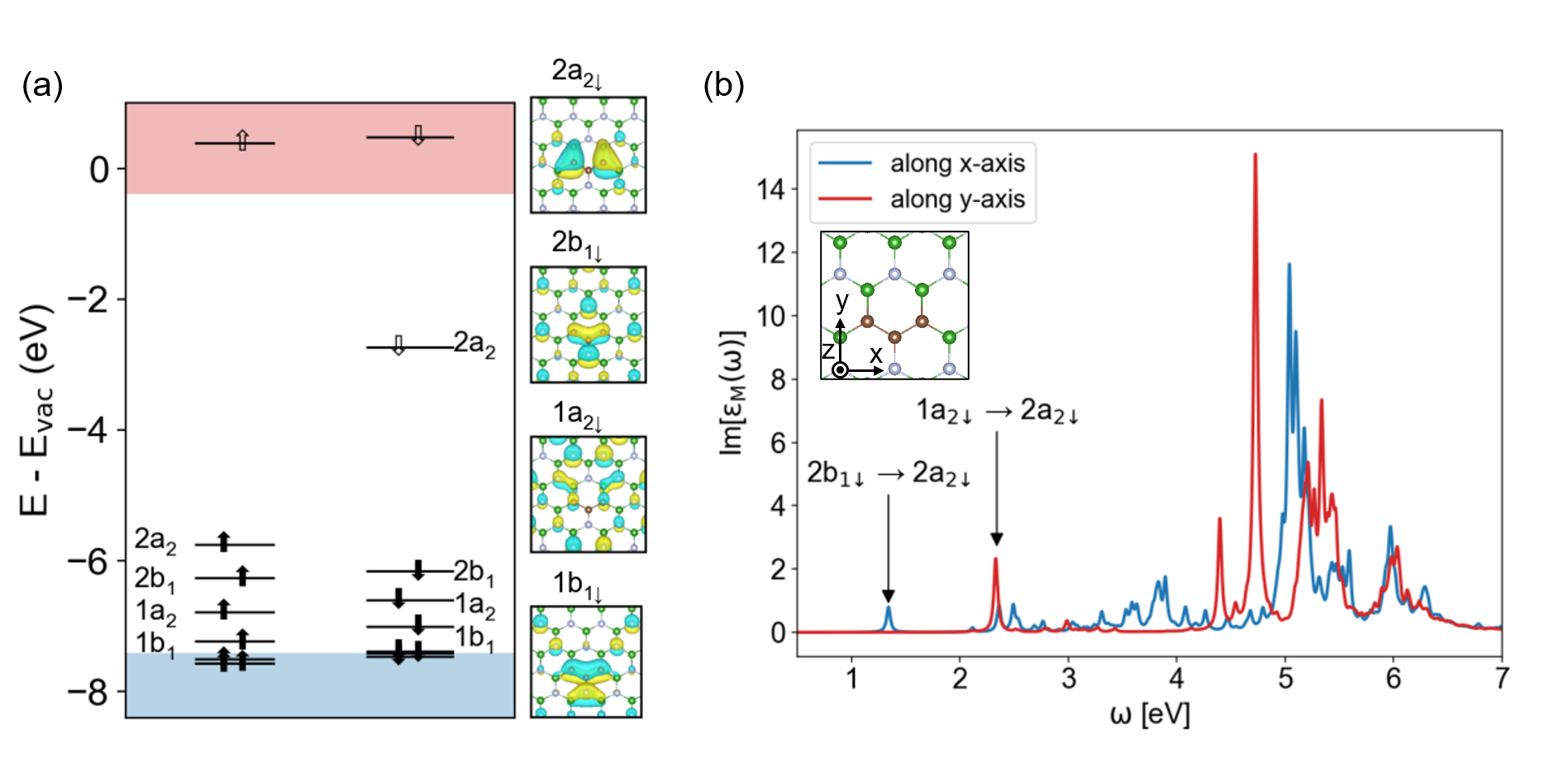}
    \caption{(a) Single-particle diagram of the ground state $\gs$ at the level of $\gw$@PBE, and (b) $\gwbse$@PBE of $\ctcn$. VBM and CBM are -7.431 eV and -0.400 eV, respectively~\cite{smart2018}. The defect states in the band gap are denoted by the irreducible representations of $C_{2v}$ symmetry group based on corresponding wavefunction symmetry. The isosurface of the wavefunctions (PBE) is $3\%$ of the maximum. In the $\gwbse$@PBE spectra, the absorption peaks are labeled by the corresponding intra-defect transitions. x and y are the in-plane directions that are perpendicular and parallel to the $C_2$ axis, respectively, and z is the out-of-plane direction. The spectral broadening is 0.02 eV. 
    }
    \label{fig:sp_diag_and_bse}
\end{figure*}

The single-particle diagram in FIG.~\ref{fig:sp_diag_and_bse}(a) 
shows the defect-related electronic energy levels with their wavefunctions, and the host monolayer hBN band edges at the GW approximation, referenced to the vacuum level.
One unoccupied defect state ($2a_2$) and three occupied defect states ($2b_1$, $1a_2$, $1b_1$) in the spin-down channel can lead to intra-defect transitions via radiative recombination or nonradiative recombination. To obtain optical transitions (or two-particle excitations), we calculate the absorption spectra by solving the BSE, which uses GW quasiparticle energies as inputs and  includes excitonic effect. Details of numerical convergence tests can be found in SI Section \rn2. Here only spin-conserved transitions are considered. We then calculate the radiative lifetimes of the intra-defect recombination via Fermi's golden rule by the Eq.~\ref{eq:rad} (with dielectric constant equal to unity for monolayer 2D systems),
\begin{equation}
\tau^{R} =\frac{3c^3}{4E_0^3\mu_{e-h}^2}\label{eq:rad},
\end{equation}
where $E_0$ is the excitation energy, and $\mu^2_{e-h}$ is modulus square of exciton dipole moment in atomic units.
The derivation detail of this equation can be found in Ref.~\citenum{palummo2015exciton, wu2019dimensionality}. We find two strong absorption peaks due to the optically allowed intra-defect transitions in FIG.~\ref{fig:sp_diag_and_bse}(b) and summarize the optical properties in TABLE~\ref{tab:rad}. In particular, only the transition $\mathrm{2a_{2\downarrow}} \rightarrow \mathrm{1a_{2\downarrow}}$ shows a relatively short radiative lifetime that possibly falls into the range of experimental values. The other transition has an order of magnitude longer radiative lifetime. Therefore, we mainly focus on the former transition. 
Its shorter lifetime (i.e. 51.9 ns) is because of stronger oscillator strength and higher excitation energy (shown in Table~\ref{tab:rad}). Note that the excitonic effect at the defect is comparable or stronger than its host materials~\cite{guo2021substrate, Attaccalite2011}, e.g. over one eV exciton binding energy $E_b$ for both transitions in Table II.
%
The exciton wavefunction in SI FIG. S3 shows that the exciton is localized and bound to the carbon defect $\ctcn$ and a few neighboring BN atoms, consistent with its large exciton binding energy. In order to compare with the experimental PL lifetimes in TABLE~\ref{tab:exp}, we also need to compute nonradiative lifetime as follows.

\begin{table}[H]
    \centering
    \caption{Radiative recombination of $\ctcn$. Excitation energy ($E_0$), square modulus of dipole moment ($\mu_{e-h}^2$), radiative lifetime ($\tau^{\mathrm{R}}$), and exciton binding energy ($E_b$) of the $\ctcn$ defect for the two transitions that are prominent in the optical excitations below the optical gap.}
    \begin{ruledtabular}
    \begin{tabular}{ccccc}
        Transition                            &  $E_0$ (eV)  &  $\mu_{e-h}^2$ (a.u.)  &  $\tau^{\mathrm{R}}$ (ns)  &  $E_b$ (eV)\\
        \hline
        $\mathrm{2a_{2\downarrow}} \rightarrow \mathrm{2b_{1\downarrow}}$ &  1.34       &  $4.90*10^{-1}$        &  $7.95*10^{2}$            &  2.21\\
        $\mathrm{2a_{2\downarrow}} \rightarrow \mathrm{1a_{2\downarrow}}$ &  2.33       &  1.43                  &  $5.19*10^{1}$           &  1.76\\
    \end{tabular}
    \end{ruledtabular}
    \label{tab:rad}
\end{table}

\subsection{Nonradiative Recombination}
Nonradiative lifetime ($\tau^{NR}$) is a measure of how fast the nonradiative recombination happens between the final state $\ket{f}$ and initial state $\ket{i}$. The phonon-assisted nonradiative recombination is influenced by several factors and also evaluated via Fermi's golden rule as below, 
\begin{align}
    \frac{1}{\tau^{NR}_{if}} &= \frac{2\pi}{\hbar}g\sum_{n,m}p_{in}|\mel{fm}{H^{e-ph}}{in}|^2\delta(E_{fm}-E_{in}) \nonumber\\
    \label{eq:NR}
\end{align}
where $H^{e-ph}$ is the electron-phonon coupling Hamiltonian, $g$ is the degeneracy factor of the final state that depends on the number of equivalent atomic configurations, and $p_{in}$ is the occupation number of the vibronic state $\ket{in}$ following Boltzmann distribution.

Under the static coupling approximation with one-dimensional (1D) phonon approximation~\cite{nonrad_alkauskas2014first, wu2019carrier}, we can rewrite Eq.~\ref{eq:NR} as
\begin{align}
\frac{1}{\tau^{NR}_{if}}
    &= \frac{2\pi}{\hbar}g|W_{if}|^2X_{if}(T) \label{eq:nonrad_wx}\\
    W_{if} &= \mel{\psi_i(\mathbf{r, R})}{\frac{\partial H}{\partial Q}}{\psi_f(\mathbf{r, R})}|_{\mathbf{R=R_a}} \label{eq:wif}\\
    X_{if} &= \sum_{n,m}p_{in}|\mel{\phi_{fm}(\mathbf{R})}{Q-Q_a}{\phi_{in}(\mathbf{R})}|^2 \nonumber\\
    &\times\delta(m\hbar\omega_f-n\hbar\omega_i+\Delta E_{if}). \label{eq:xif}
\end{align}
  Eq.~\ref{eq:nonrad_wx} is separated into the electronic term ($W_{if}$) that depends on the electronic wavefunction ($\psi$) overlap, and the phonon term ($X_{if}$) that describes the strength of phonon contribution. The phonon term includes the energy conservation between initial and final vibronic states with vibrational frequencies of $\omega_i$ and $\omega_f$, and $\phi$ is the phonon wavefunction. The detailed derivation can be found in Ref.~\citenum{nonrad_alkauskas2014first, wu2019carrier}.

We summarize the nonradiative recombination lifetime of the intra-defect transitions, along with ZPL, HR factor of the final state ($S_f(1D)$) in TABLE~\ref{tab:nonrad}. Note that the HR factor from the nonradiative recombination calculation is with 1D effective phonon approximation. Its comparison with the full-phonon HR factor ($S$, including all phonon eigenmodes) can be found in SI Section \rn5. The calculated 1D HR factor is close to the full-phonon HR factor as can be seen by comparing TABLE~\ref{tab:nonrad} (1D) with FIG.~\ref{fig:multiplet_c2cn} (full phonon). ZPL is evaluated by two methods: one is using the vertical neutral excitation energy obtained from BSE with subtracting the Frank-Condon shift in the excited state (ZPL(BSE)), and another is from constrained DFT (CDFT) at the PBE level with geometry optimization (ZPL(CDFT)). The ZPL from two methods has 0.1 and 0.4 eV energy difference for the first two transitions, respectively. We used the CDFT in the nonradiative lifetime calculation to be consistent with other quantities in the equation.
The nonradiative recombination of $\est \rightarrow \gs$ is fast due to the following reason. The electronic term $W_{if}$ of $\est \rightarrow \gs$ is large because it is symmetry-allowed, about four orders of magnitude larger than the other two transitions.
%
%
Since $\tau^{NR}$ is inversely proportional to the square of $W_{if}$  according to Eq.~\ref{eq:nonrad_wx}, difference of $W_{if}$ dominates over phonon contribution in $X_{if}$. 
The nonradiative lifetime of $\est \rightarrow \gs$ (2.29 ns) is several orders shorter than other two transitions.

\begin{table*}
    \centering
    \caption{Properties of nonradiative recombination of the intra-defect transitions. The transitions are denoted using the multielectron wavefunction notations. The ZPL(BSE) are evaluated by subtracting the Franck-Condon shift ($E_{\mathrm{FC}}$) from the excitation energies from $\gwbse$ calculations. The ZPL(CDFT) are obtained by constrained DFT calculations at PBE level, which are used as the energy input for the nonradiative lifetime. $\Delta Q$ is the nuclear coordinate change between the initial and final states. $\hbar\omega_f$ is the phonon energy of the final state. $S_f(1D)$ is the HR factor with 1D effective phonon approximation. $W_{if}$ and $X_{if}$ are the electronic and phonon terms, respectively. The nonradiative lifetimes are calculated with $6\times6$ supercell at 300 K at PBE level. Here, $\esb \rightarrow \gs$ is related to the transition $\mathrm{2a_{2\downarrow}} \rightarrow \mathrm{2b_{1\downarrow}}$ in the single particle picture; transition $\est \rightarrow \gs$ is related to the transition $\mathrm{2a_{2\downarrow}} \rightarrow \mathrm{1a_{2\downarrow}}$; transition $\est \rightarrow \esb$ is related to the transition $\mathrm{2b_{1\downarrow}} \rightarrow \mathrm{1a_{2\downarrow}}$. } 
    \begin{ruledtabular}
    \begin{tabular}{cccccccccc}
        Transition & $E_{FC}$ &  ZPL(BSE) & ZPL(CDFT) & $\Delta Q$ & $\hbar\omega_f$ & $S_f(1D)$ & $W_{if}$ &  $X_{if}$ &  $\tau^{\mathrm{NR}}$\\
                   &  (eV) & (eV) & (eV)& (amu$^{1/2}\si{\angstrom}$) & (meV) & & (eV/(amu$^{1/2}\si{\angstrom}$))  &  (amu$\cdot\si{\angstrom}^2$/eV)  &  \\
        \hline
        $\esb \rightarrow \gs$   & 0.13 & 1.21 & 1.16 & 0.26 & 122 & 0.95 & $1.34*10^{-4}$ &  $7.17*10^{-6}$ &  ${\sim}1$ ms \\
        $\est \rightarrow \gs$   & 0.20 & 2.13 & 1.70 & 0.27 & 145 & 1.26 & $3.13*10^{-1}$ &  $4.69*10^{-7}$ &  2.29 ns \\
        $\est \rightarrow \esb$  & 0.01  & -- & 0.55 & 0.19 & 128 & 0.67 & $5.58*10^{-5}$ & $1.05*10^{-2}$ & 3.20 $\mu$s \\
    \end{tabular}
    \end{ruledtabular}
    \label{tab:nonrad}
\end{table*}

\subsection{PL Lifetime and Quantum Yield}
\begin{figure}
    \centering
    \includegraphics[width=0.48\textwidth]{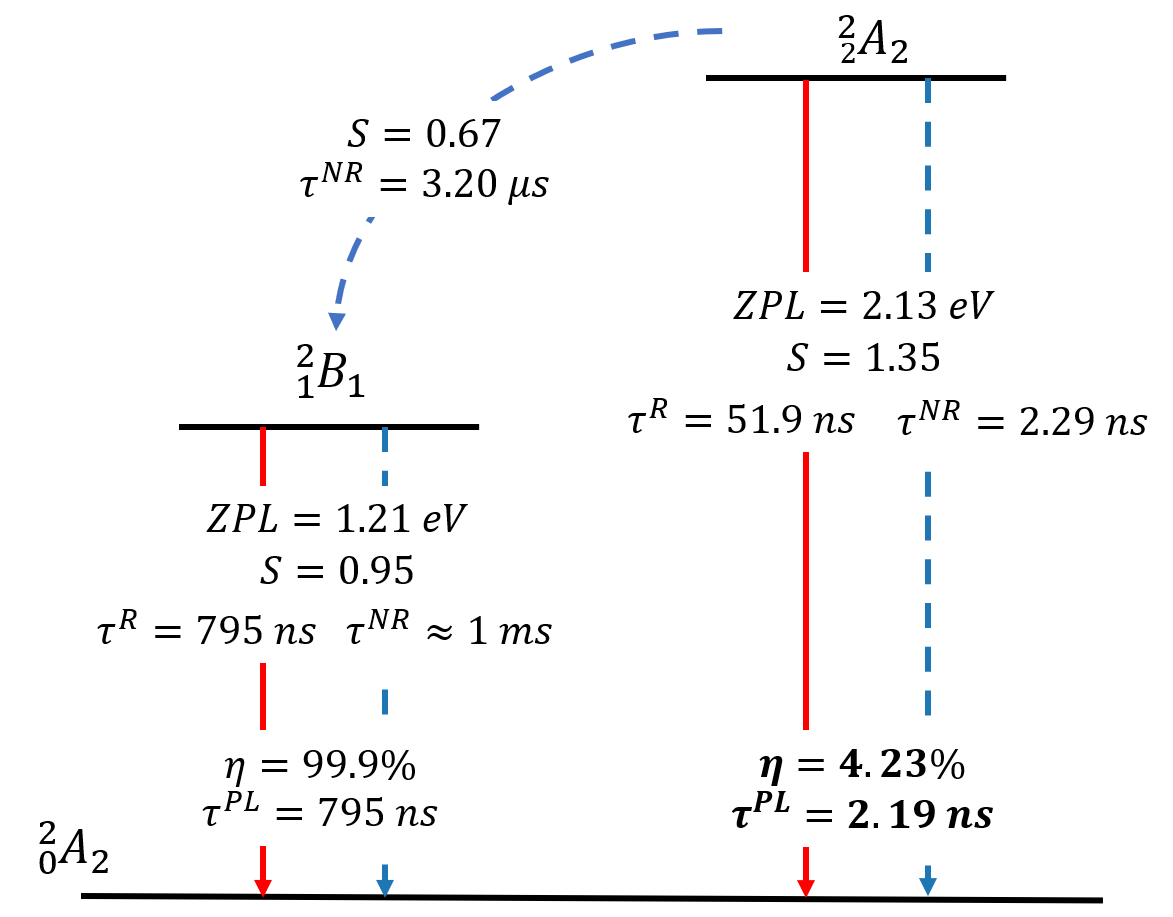}
    \caption{Multiplet structure and important physical parameters of $\ctcn$ in hBN. 
    %
    %
    $\gs$ is the ground state, and $\esb$ and $\est$ are the excited states. The red solid lines denote radiative recombination and the blue dashed lines denote nonradiative recombination. $S$ is the HR factor with full phonon calculations. $\tau^R$ and $\tau^{NR}$ are the radiative lifetime and nonradiative lifetime, respectively. $\eta$ and $\tau^{PL}$ are the QY and PL lifetime, respectively. 
    }
    \label{fig:multiplet_c2cn}
\end{figure}

Next we will compute PL lifetime and quantum yield~\cite{turiansky2021impact} using radiative and nonradiative recombination lifetime from TABLE~\ref{tab:rad} and TABLE~\ref{tab:nonrad}, respectively. We list them along with the full-phonon HR factors calculated with the method in Ref.~\citenum{pl_alkauskas2014first}, and quantum yield ($\eta$) computed by Eq.~\ref{eq:qy}~\cite{novotny2012principles} in the multiplet diagram FIG.~\ref{fig:multiplet_c2cn}.  
\begin{align}
    \eta &= \frac{1}{\tau^{\mathrm{R}}} / (\frac{1}{\tau^{\mathrm{R}}} + \sum_i\frac{1}{\tau_i^{\mathrm{NR}}}), \label{eq:qy}
\end{align}
where $i$ denotes the $i^{th}$ nonradiative pathway for the transition from the initial state.
In addition, we calculate the PL lifetime which is the inverse of the total recombination rate by Eq.~\ref{eq:tau_pl} 
\begin{equation}
    \tau^{\mathrm{PL}} = 1/(\frac{1}{\tau^{\mathrm{R}}} + \sum_i\frac{1}{\tau_i^{\mathrm{NR}}}) = \tau^{\mathrm{R}} * \eta \label{eq:tau_pl}.
\end{equation}

For transition $\est \rightarrow \gs$, the nonradiative lifetime is nearly ten times shorter than the radiative lifetime. By including the nonradiative recombination $\est \rightarrow \esb$, the overall recombination leads to the quantum yield of $4.23\%$, about $2\%$ lower than the experimental quantum yield range $6{\sim}12\%$~\cite{qyli2019near}. More importantly, the PL lifetime of transition $\est \rightarrow \gs$ from Eq.~\ref{eq:tau_pl} is calculated to be 2.19 ns, exactly within the experimental range of PL lifetimes (TABLE~\ref{tab:exp} and FIG.~\ref{fig:theory_vs_exp}(a)). In addition, the calculated full-phonon HR factor of this transition is 1.35, indicating weak electron-phonon interaction and good agreement with the experimental HR factor in TABLE~\ref{tab:exp} and FIG.~\ref{fig:theory_vs_exp}(b).

\subsection{Phonon Modes of $\ctcn$ in hBN}
\begin{figure}
    \centering
    \includegraphics[width=0.45\textwidth]{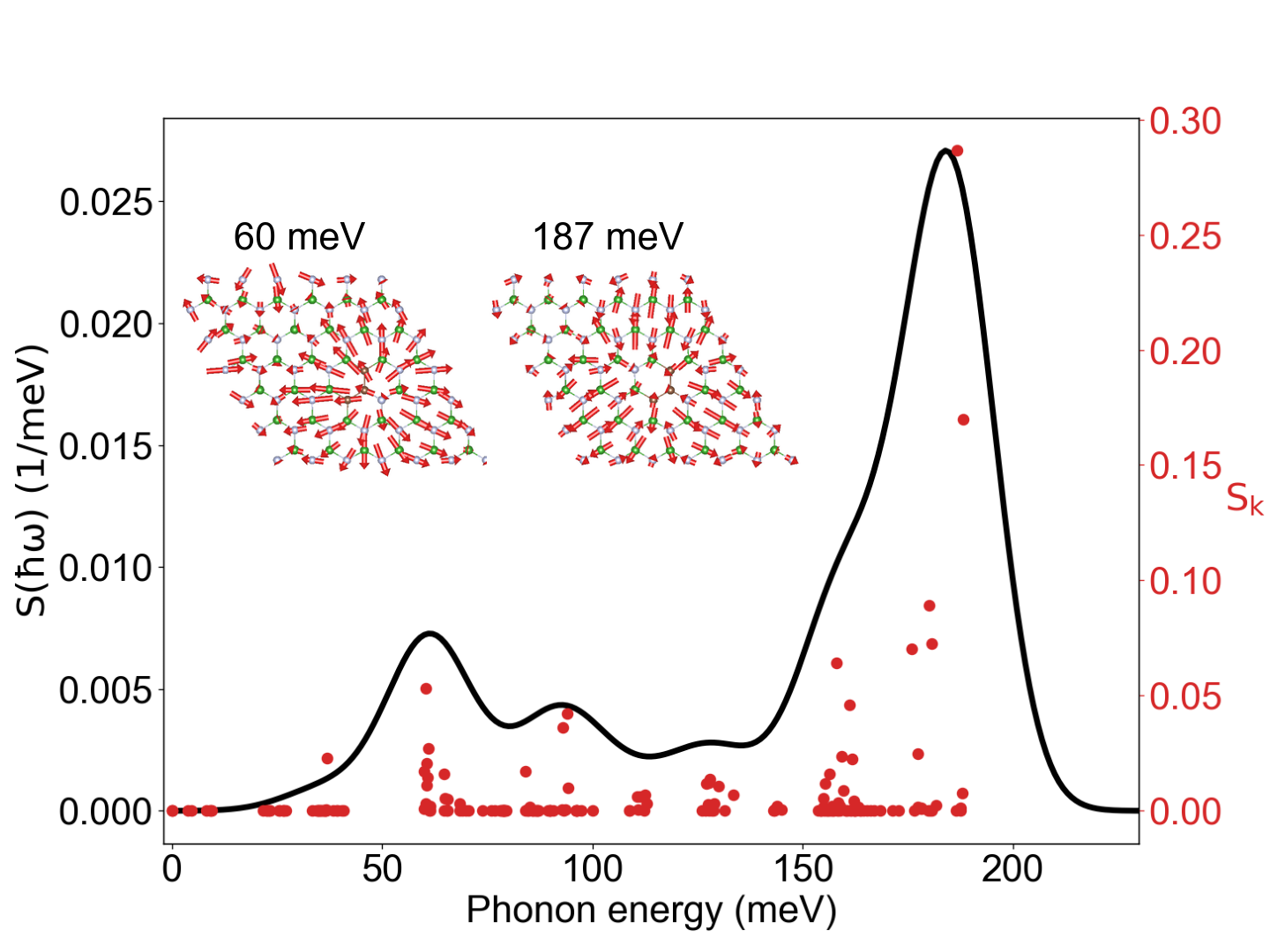}
    \caption{Spectral function that shows the distribution of phonon modes and the contribution of phonon modes to the electron-phonon interaction of $\ctcn$ defect in hBN. The left vertical axis and black solid line are for the spectral function, and the right vertical axis and red dots are for the partial HR factor as a function of phonon energy. The inset figures are the low energy phonon mode (60 meV) and dominant phonon mode (187 meV) of $\ctcn$ for transition $\est \rightarrow \gs$.The red arrows show the atom displacement of the corresponding phonon modes.}
    \label{fig:phonon}
\end{figure}

To get more insights into electron-phonon coupling of $\ctcn$ in hBN, we calculate the spectral function ($S(\hbar\omega)$) of $\est \rightarrow \gs$ with the partial HR factor ($S_k$) of phonon mode $k$ as shown in FIG.~\ref{fig:phonon}. The phonon modes spread over the range $0{\sim}190$ meV, and the phonon modes around 180 meV have the largest contribution to the spectral function. In particular, the 187 meV phonon mode has the largest partial HR factor among these modes. By visualizing this phonon mode shown by the inset in FIG.~\ref{fig:phonon}, we find it to be an out-of-phase phonon mode. The atoms involved in this phonon mode include not only defect-related atoms but also many B and N atoms around and away from the defect. Hence, the 187 meV phonon mode is quasi-local. To quantitatively characterize the localization of dominant phonon modes, we estimate the number of atoms involved in the vibration of a phonon mode, by projection of inverse participation ratio (IPR) on this phonon mode $k$~\cite{pl_alkauskas2014first},
\begin{align}
      \mathrm{IPR_k} &= \frac{1}{\sum_{\alpha} (\sum_{i}\Delta r^2_{k;\alpha,i})^2},  
\end{align}
where $\Delta r_{k;\alpha,i}$ is 
a normalized vector that describes the displacement of the atom $\alpha$ along the direction $i$ in the phonon mode $k$ between the intial and final states. We then estimate the localization of the phonon mode by the localization ratio $\beta$,
\begin{align}
    \beta_k &= \mathrm{NAT/IPR_k},
\end{align}
where NAT stands for the total number of atoms in the supercell. The larger the localization ratio is, the more localized the phonon mode is. We find that the IPR of the 187 meV phonon mode is 46 for the 6x6 supercell calculation, and the localization ratio $\beta$ is ${\sim}2$. The IPR of this mode increases as a function of supercell sizes, but the localization ratio keeps constant, as can be seen in the SI TABLE S4. This again reflects the nature of this mode being a mixture of defect and bulk phonons or a ``quasi-local'' mode. 
Besides the 187 meV phonon, the 60 meV phonon is pronounced in the low-energy range of the spectral function, 
as an in-phase phonon mode (see the inset in FIG.~\ref{fig:phonon}). It also has a large IPR of 48 and $\beta$ of ${\sim}2$. 
Comparing the calculated phonon mode energies with the experimental ones for 2eV single-photon emitters~\cite{zpl_raman_pse_tau_spe_grosso2017tunable, preuss2021assembly}, the 187 meV phonon mode is in good agreement with the ${\sim}180$ meV longitudinal optical phonon that leads to the pronounced PSB.

\subsection{Comparison between $\ctcn$ defect and Experiments}
\begin{figure*}
    \centering
    \includegraphics[width=0.8\textwidth]{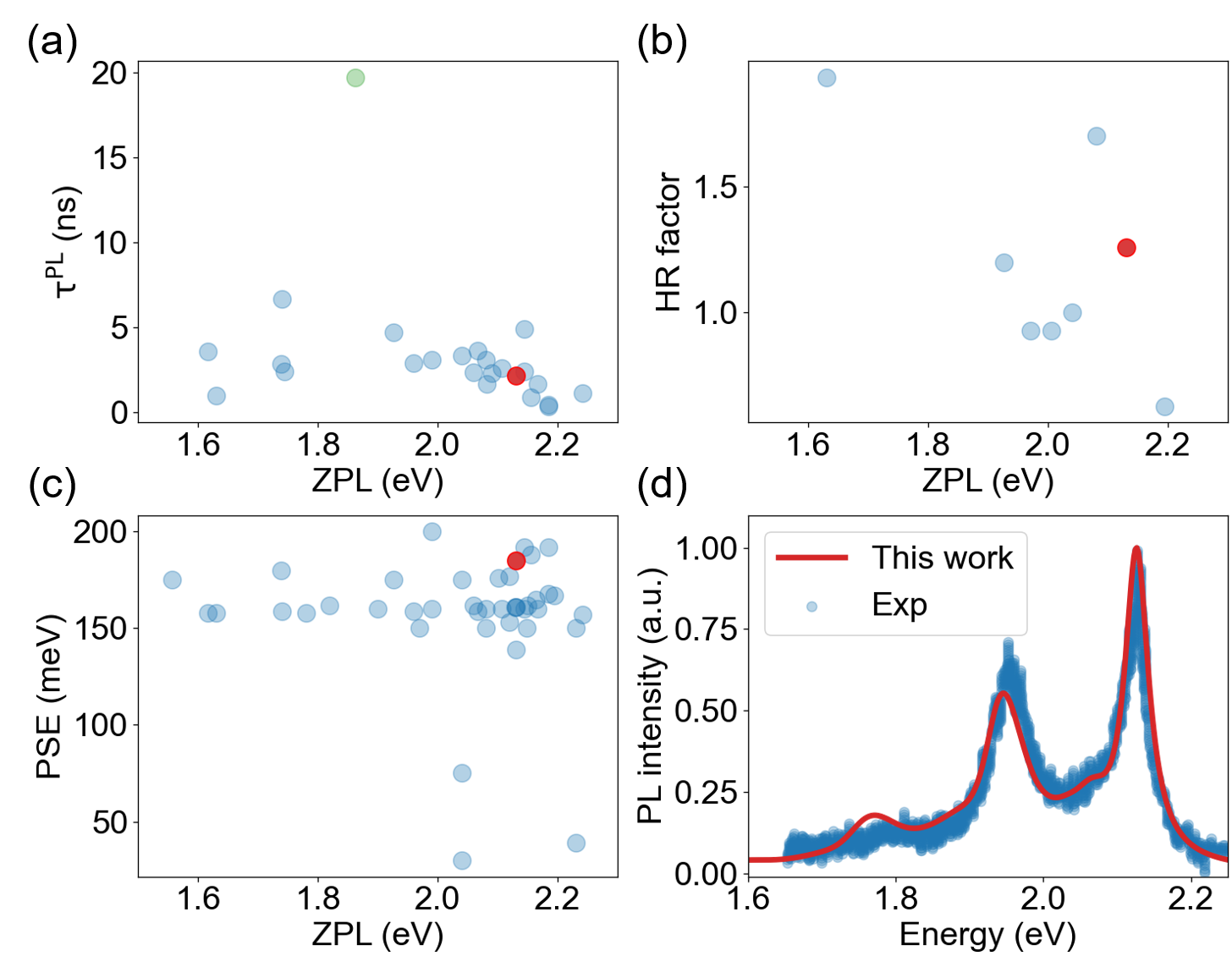}
    \caption{The calculated properties of $\ctcn$ along with the experimental values which are summarized in TABLE.~\ref{tab:exp}. (a) PL lifetime vs ZPL; (b) HR factor vs ZPL; (c) PSE vs ZPL; (d) comparison between the theoretical PL spectrum (this work) and the experimental PL spectrum (Exp)~\cite{zpl_raman_pse_spe_linpol_spinpol_mendelson2020identifying}. In (a), (b) and (c), the blue dots are for the experimental values of ${\sim}2$ eV single-photon emitters, the green dot is for the experimental PL lifetime of ${\sim}20$ ns, and the red dots are for the calculated values. The calculated ZPL here is from $\gwbse$@PBE calculation. The comparison shows that ZPL, PL lifetime, HR factor and PSE are all in the experimental range. The calculated PL spectrum also matches well with the experimental PL of the ${\sim}2$ eV single-photon emission~\cite{zpl_raman_pse_spe_linpol_spinpol_mendelson2020identifying}. 
    }
    \label{fig:theory_vs_exp}
\end{figure*}

To further validate that $\ctcn$ is a possible candidate of experimentally observed single-photon emitters, we calculate the PL of $\ctcn$ for transition $\est \rightarrow \gs$ and compare it with the PL spectra of ${\sim}2$ eV SPEs~\cite{zpl_raman_pse_spe_linpol_spinpol_mendelson2020identifying} in FIG.~\ref{fig:theory_vs_exp}. 
The details of the PL calculation can be found in SI Section \rn5. The calculated PL shows similar PSB peaks to the experimental PL~\cite{zpl_raman_pse_spe_linpol_spinpol_mendelson2020identifying}, as can be seen in FIG.~\ref{fig:theory_vs_exp}(d). A weak PSB peak next to the first PSB peak can be seen in both calculated and experimental PL spectra. This peak has been found to be more visible when the brightness is enhanced in an as-prepared array~\cite{preuss2021assembly}. In addition, phonon-sideband energy can be a measure of the averaged phonon energy, found to be 180 meV, within the experimental range in TABLE~\ref{tab:exp} and FIG.~\ref{fig:theory_vs_exp}(c). The excellent agreement between the calculated and experimental physical parameters including ZPL, PL lifetime, HR factor, PSE and PL spectrum strongly suggest that the $\ctcn$ defect is one possible source of ${\sim}2$ eV single-photon emission.
Finally, we note that the variation in experimental results shown in TABLE~\ref{tab:exp} can be from several sources of single-phonon emitters and possible internal strains due to material preparation. Further defect identification including spin related properties can be through calculations of optically detected magnetic resonance (ODMR), which measures the spin-dependent optical contrast and is directly related to the information readout of spin qubits~\cite{auburger2021towards}.

\section{Conclusions}
In conclusion, we investigated the possibilities of carbon defects for 2 eV single photon emission in hBN, by comparing key physical properties from first-principles calculations with experiments, including ZPL, PL lifetime and lineshape, as well as HR factor.
We showcase the importance of considering multiple key signatures including both static and dynamical properties when identifying defects through theory and experimental comparison. 
We found $\ctcn$ has the best agreement with experiments on all concerned properties. In particular, we show the radiative lifetime can be an order longer than the experimentally observed PL lifetime; but after including nonradiative processes, the agreement with experimental PL lifetime is excellent. 
We show the electron-phonon coupling in $\ctcn$ is moderate with small HR factor and the dominant phonon mode at ${\sim}$180 meV is quasi-local with significant participation of bulk atoms. 
%
%
Our work provides insight into the ${\sim}2$ eV single-photon emission from the theoretical perspective, which is important for unraveling the unknown chemical nature of defects in hBN, manipulating defects and developing quantum applications.

\begin{acknowledgments}
We acknowledge the support by the National Science Foundation under grant no. DMR-1760260.
This research used resources of the Scientific Data and Computing center, a component of the Computational Science Initiative, at Brookhaven National Laboratory under Contract No. DE-SC0012704,
the lux supercomputer at UC Santa Cruz, funded by NSF MRI grant AST 1828315,
the National Energy Research Scientific Computing Center (NERSC) a U.S. Department of Energy Office of Science User Facility operated under Contract No. DE-AC02-05CH11231,
the Extreme Science and Engineering Discovery Environment (XSEDE) which is supported by National Science Foundation Grant No. ACI-1548562 \cite{xsede}.
\end{acknowledgments}

\bibliography{ref}

\begin{thebibliography}{64}%
\makeatletter
\providecommand \@ifxundefined [1]{%
 \@ifx{#1\undefined}
}%
\providecommand \@ifnum [1]{%
 \ifnum #1\expandafter \@firstoftwo
 \else \expandafter \@secondoftwo
 \fi
}%
\providecommand \@ifx [1]{%
 \ifx #1\expandafter \@firstoftwo
 \else \expandafter \@secondoftwo
 \fi
}%
\providecommand \natexlab [1]{#1}%
\providecommand \enquote  [1]{``#1''}%
\providecommand \bibnamefont  [1]{#1}%
\providecommand \bibfnamefont [1]{#1}%
\providecommand \citenamefont [1]{#1}%
\providecommand \href@noop [0]{\@secondoftwo}%
\providecommand \href [0]{\begingroup \@sanitize@url \@href}%
\providecommand \@href[1]{\@@startlink{#1}\@@href}%
\providecommand \@@href[1]{\endgroup#1\@@endlink}%
\providecommand \@sanitize@url [0]{\catcode `\\12\catcode `\$12\catcode
  `\&12\catcode `\#12\catcode `\^12\catcode `\_12\catcode `\%12\relax}%
\providecommand \@@startlink[1]{}%
\providecommand \@@endlink[0]{}%
\providecommand \url  [0]{\begingroup\@sanitize@url \@url }%
\providecommand \@url [1]{\endgroup\@href {#1}{\urlprefix }}%
\providecommand \urlprefix  [0]{URL }%
\providecommand \Eprint [0]{\href }%
\providecommand \doibase [0]{https://doi.org/}%
\providecommand \selectlanguage [0]{\@gobble}%
\providecommand \bibinfo  [0]{\@secondoftwo}%
\providecommand \bibfield  [0]{\@secondoftwo}%
\providecommand \translation [1]{[#1]}%
\providecommand \BibitemOpen [0]{}%
\providecommand \bibitemStop [0]{}%
\providecommand \bibitemNoStop [0]{.\EOS\space}%
\providecommand \EOS [0]{\spacefactor3000\relax}%
\providecommand \BibitemShut  [1]{\csname bibitem#1\endcsname}%
\let\auto@bib@innerbib\@empty
\bibitem [{\citenamefont {Wang}\ \emph {et~al.}(2020)\citenamefont {Wang},
  \citenamefont {Sciarrino}, \citenamefont {Laing},\ and\ \citenamefont
  {Thompson}}]{wang2020integrated}%
  \BibitemOpen
  \bibfield  {author} {\bibinfo {author} {\bibfnamefont {J.}~\bibnamefont
  {Wang}}, \bibinfo {author} {\bibfnamefont {F.}~\bibnamefont {Sciarrino}},
  \bibinfo {author} {\bibfnamefont {A.}~\bibnamefont {Laing}},\ and\ \bibinfo
  {author} {\bibfnamefont {M.~G.}\ \bibnamefont {Thompson}},\ }\bibfield
  {title} {\bibinfo {title} {Integrated photonic quantum technologies},\
  }\href@noop {} {\bibfield  {journal} {\bibinfo  {journal} {Nat. Photonics}\
  }\textbf {\bibinfo {volume} {14}},\ \bibinfo {pages} {273} (\bibinfo {year}
  {2020})}\BibitemShut {NoStop}%
\bibitem [{\citenamefont {Wang}\ \emph {et~al.}(2019)\citenamefont {Wang},
  \citenamefont {Qin}, \citenamefont {Ding}, \citenamefont {Chen},
  \citenamefont {Chen}, \citenamefont {You}, \citenamefont {He}, \citenamefont
  {Jiang}, \citenamefont {You}, \citenamefont {Wang} \emph
  {et~al.}}]{wang2019boson}%
  \BibitemOpen
  \bibfield  {author} {\bibinfo {author} {\bibfnamefont {H.}~\bibnamefont
  {Wang}}, \bibinfo {author} {\bibfnamefont {J.}~\bibnamefont {Qin}}, \bibinfo
  {author} {\bibfnamefont {X.}~\bibnamefont {Ding}}, \bibinfo {author}
  {\bibfnamefont {M.-C.}\ \bibnamefont {Chen}}, \bibinfo {author}
  {\bibfnamefont {S.}~\bibnamefont {Chen}}, \bibinfo {author} {\bibfnamefont
  {X.}~\bibnamefont {You}}, \bibinfo {author} {\bibfnamefont {Y.-M.}\
  \bibnamefont {He}}, \bibinfo {author} {\bibfnamefont {X.}~\bibnamefont
  {Jiang}}, \bibinfo {author} {\bibfnamefont {L.}~\bibnamefont {You}}, \bibinfo
  {author} {\bibfnamefont {Z.}~\bibnamefont {Wang}}, \emph {et~al.},\
  }\bibfield  {title} {\bibinfo {title} {Boson sampling with 20 input photons
  and a 60-mode interferometer in a 1 0 14-dimensional hilbert space},\
  }\href@noop {} {\bibfield  {journal} {\bibinfo  {journal} {Phys. Rev. Lett.}\
  }\textbf {\bibinfo {volume} {123}},\ \bibinfo {pages} {250503} (\bibinfo
  {year} {2019})}\BibitemShut {NoStop}%
\bibitem [{\citenamefont {Barrett}\ \emph {et~al.}(2005)\citenamefont
  {Barrett}, \citenamefont {Hardy},\ and\ \citenamefont
  {Kent}}]{barrett2005no}%
  \BibitemOpen
  \bibfield  {author} {\bibinfo {author} {\bibfnamefont {J.}~\bibnamefont
  {Barrett}}, \bibinfo {author} {\bibfnamefont {L.}~\bibnamefont {Hardy}},\
  and\ \bibinfo {author} {\bibfnamefont {A.}~\bibnamefont {Kent}},\ }\bibfield
  {title} {\bibinfo {title} {No signaling and quantum key distribution},\
  }\href@noop {} {\bibfield  {journal} {\bibinfo  {journal} {Phys. Rev. Lett.}\
  }\textbf {\bibinfo {volume} {95}},\ \bibinfo {pages} {010503} (\bibinfo
  {year} {2005})}\BibitemShut {NoStop}%
\bibitem [{\citenamefont {Cassabois}\ \emph {et~al.}(2016)\citenamefont
  {Cassabois}, \citenamefont {Valvin},\ and\ \citenamefont
  {Gil}}]{cassabois2016hexagonal}%
  \BibitemOpen
  \bibfield  {author} {\bibinfo {author} {\bibfnamefont {G.}~\bibnamefont
  {Cassabois}}, \bibinfo {author} {\bibfnamefont {P.}~\bibnamefont {Valvin}},\
  and\ \bibinfo {author} {\bibfnamefont {B.}~\bibnamefont {Gil}},\ }\bibfield
  {title} {\bibinfo {title} {Hexagonal boron nitride is an indirect bandgap
  semiconductor},\ }\href@noop {} {\bibfield  {journal} {\bibinfo  {journal}
  {Nat. Photonics}\ }\textbf {\bibinfo {volume} {10}},\ \bibinfo {pages} {262}
  (\bibinfo {year} {2016})}\BibitemShut {NoStop}%
\bibitem [{\citenamefont {Elias}\ \emph {et~al.}(2019)\citenamefont {Elias},
  \citenamefont {Valvin}, \citenamefont {Pelini}, \citenamefont {Summerfield},
  \citenamefont {Mellor}, \citenamefont {Cheng}, \citenamefont {Eaves},
  \citenamefont {Foxon}, \citenamefont {Beton}, \citenamefont {Novikov} \emph
  {et~al.}}]{elias2019direct}%
  \BibitemOpen
  \bibfield  {author} {\bibinfo {author} {\bibfnamefont {C.}~\bibnamefont
  {Elias}}, \bibinfo {author} {\bibfnamefont {P.}~\bibnamefont {Valvin}},
  \bibinfo {author} {\bibfnamefont {T.}~\bibnamefont {Pelini}}, \bibinfo
  {author} {\bibfnamefont {A.}~\bibnamefont {Summerfield}}, \bibinfo {author}
  {\bibfnamefont {C.}~\bibnamefont {Mellor}}, \bibinfo {author} {\bibfnamefont
  {T.}~\bibnamefont {Cheng}}, \bibinfo {author} {\bibfnamefont
  {L.}~\bibnamefont {Eaves}}, \bibinfo {author} {\bibfnamefont
  {C.}~\bibnamefont {Foxon}}, \bibinfo {author} {\bibfnamefont
  {P.}~\bibnamefont {Beton}}, \bibinfo {author} {\bibfnamefont
  {S.}~\bibnamefont {Novikov}}, \emph {et~al.},\ }\bibfield  {title} {\bibinfo
  {title} {Direct band-gap crossover in epitaxial monolayer boron nitride},\
  }\href@noop {} {\bibfield  {journal} {\bibinfo  {journal} {Nat. Commun.}\
  }\textbf {\bibinfo {volume} {10}},\ \bibinfo {pages} {1} (\bibinfo {year}
  {2019})}\BibitemShut {NoStop}%
\bibitem [{\citenamefont {Fischer}\ \emph {et~al.}(2021)\citenamefont
  {Fischer}, \citenamefont {Caridad}, \citenamefont {Sajid}, \citenamefont
  {Ghaderzadeh}, \citenamefont {Ghorbani-Asl}, \citenamefont {Gammelgaard},
  \citenamefont {B{\o}ggild}, \citenamefont {Thygesen}, \citenamefont
  {Krasheninnikov}, \citenamefont {Xiao} \emph
  {et~al.}}]{fischer2021controlled}%
  \BibitemOpen
  \bibfield  {author} {\bibinfo {author} {\bibfnamefont {M.}~\bibnamefont
  {Fischer}}, \bibinfo {author} {\bibfnamefont {J.}~\bibnamefont {Caridad}},
  \bibinfo {author} {\bibfnamefont {A.}~\bibnamefont {Sajid}}, \bibinfo
  {author} {\bibfnamefont {S.}~\bibnamefont {Ghaderzadeh}}, \bibinfo {author}
  {\bibfnamefont {M.}~\bibnamefont {Ghorbani-Asl}}, \bibinfo {author}
  {\bibfnamefont {L.}~\bibnamefont {Gammelgaard}}, \bibinfo {author}
  {\bibfnamefont {P.}~\bibnamefont {B{\o}ggild}}, \bibinfo {author}
  {\bibfnamefont {K.~S.}\ \bibnamefont {Thygesen}}, \bibinfo {author}
  {\bibfnamefont {A.}~\bibnamefont {Krasheninnikov}}, \bibinfo {author}
  {\bibfnamefont {S.}~\bibnamefont {Xiao}}, \emph {et~al.},\ }\bibfield
  {title} {\bibinfo {title} {Controlled generation of luminescent centers in
  hexagonal boron nitride by irradiation engineering},\ }\href@noop {}
  {\bibfield  {journal} {\bibinfo  {journal} {Sci. Adv.}\ }\textbf {\bibinfo
  {volume} {7}},\ \bibinfo {pages} {eabe7138} (\bibinfo {year}
  {2021})}\BibitemShut {NoStop}%
\bibitem [{\citenamefont {Grosso}\ \emph {et~al.}(2017)\citenamefont {Grosso},
  \citenamefont {Moon}, \citenamefont {Lienhard}, \citenamefont {Ali},
  \citenamefont {Efetov}, \citenamefont {Furchi}, \citenamefont
  {Jarillo-Herrero}, \citenamefont {Ford}, \citenamefont {Aharonovich},\ and\
  \citenamefont {Englund}}]{zpl_raman_pse_tau_spe_grosso2017tunable}%
  \BibitemOpen
  \bibfield  {author} {\bibinfo {author} {\bibfnamefont {G.}~\bibnamefont
  {Grosso}}, \bibinfo {author} {\bibfnamefont {H.}~\bibnamefont {Moon}},
  \bibinfo {author} {\bibfnamefont {B.}~\bibnamefont {Lienhard}}, \bibinfo
  {author} {\bibfnamefont {S.}~\bibnamefont {Ali}}, \bibinfo {author}
  {\bibfnamefont {D.~K.}\ \bibnamefont {Efetov}}, \bibinfo {author}
  {\bibfnamefont {M.~M.}\ \bibnamefont {Furchi}}, \bibinfo {author}
  {\bibfnamefont {P.}~\bibnamefont {Jarillo-Herrero}}, \bibinfo {author}
  {\bibfnamefont {M.~J.}\ \bibnamefont {Ford}}, \bibinfo {author}
  {\bibfnamefont {I.}~\bibnamefont {Aharonovich}},\ and\ \bibinfo {author}
  {\bibfnamefont {D.}~\bibnamefont {Englund}},\ }\bibfield  {title} {\bibinfo
  {title} {Tunable and high-purity room temperature single-photon emission from
  atomic defects in hexagonal boron nitride},\ }\href@noop {} {\bibfield
  {journal} {\bibinfo  {journal} {Nat. Commun.}\ }\textbf {\bibinfo {volume}
  {8}},\ \bibinfo {pages} {1} (\bibinfo {year} {2017})}\BibitemShut {NoStop}%
\bibitem [{\citenamefont {Li}\ \emph {et~al.}(2019)\citenamefont {Li},
  \citenamefont {Scully}, \citenamefont {Shayan}, \citenamefont {Luo},\ and\
  \citenamefont {Strauf}}]{qyli2019near}%
  \BibitemOpen
  \bibfield  {author} {\bibinfo {author} {\bibfnamefont {X.}~\bibnamefont
  {Li}}, \bibinfo {author} {\bibfnamefont {R.~A.}\ \bibnamefont {Scully}},
  \bibinfo {author} {\bibfnamefont {K.}~\bibnamefont {Shayan}}, \bibinfo
  {author} {\bibfnamefont {Y.}~\bibnamefont {Luo}},\ and\ \bibinfo {author}
  {\bibfnamefont {S.}~\bibnamefont {Strauf}},\ }\bibfield  {title} {\bibinfo
  {title} {Near-unity light collection efficiency from quantum emitters in
  boron nitride by coupling to metallo-dielectric antennas},\ }\href@noop {}
  {\bibfield  {journal} {\bibinfo  {journal} {ACS Nano}\ }\textbf {\bibinfo
  {volume} {13}},\ \bibinfo {pages} {6992} (\bibinfo {year}
  {2019})}\BibitemShut {NoStop}%
\bibitem [{\citenamefont {Tran}\ \emph
  {et~al.}(2016{\natexlab{a}})\citenamefont {Tran}, \citenamefont {Bray},
  \citenamefont {Ford}, \citenamefont {Toth},\ and\ \citenamefont
  {Aharonovich}}]{zpl_raman_dw_tau_qy_spe_linpol_tran2016quantummono}%
  \BibitemOpen
  \bibfield  {author} {\bibinfo {author} {\bibfnamefont {T.~T.}\ \bibnamefont
  {Tran}}, \bibinfo {author} {\bibfnamefont {K.}~\bibnamefont {Bray}}, \bibinfo
  {author} {\bibfnamefont {M.~J.}\ \bibnamefont {Ford}}, \bibinfo {author}
  {\bibfnamefont {M.}~\bibnamefont {Toth}},\ and\ \bibinfo {author}
  {\bibfnamefont {I.}~\bibnamefont {Aharonovich}},\ }\bibfield  {title}
  {\bibinfo {title} {Quantum emission from hexagonal boron nitride
  monolayers},\ }\href@noop {} {\bibfield  {journal} {\bibinfo  {journal} {Nat.
  Nanotechnol.}\ }\textbf {\bibinfo {volume} {11}},\ \bibinfo {pages} {37}
  (\bibinfo {year} {2016}{\natexlab{a}})}\BibitemShut {NoStop}%
\bibitem [{\citenamefont {Tran}\ \emph
  {et~al.}(2016{\natexlab{b}})\citenamefont {Tran}, \citenamefont {Zachreson},
  \citenamefont {Berhane}, \citenamefont {Bray}, \citenamefont {Sandstrom},
  \citenamefont {Li}, \citenamefont {Taniguchi}, \citenamefont {Watanabe},
  \citenamefont {Aharonovich},\ and\ \citenamefont
  {Toth}}]{zpl_raman_hr_tau_spe_tran2016quantum}%
  \BibitemOpen
  \bibfield  {author} {\bibinfo {author} {\bibfnamefont {T.~T.}\ \bibnamefont
  {Tran}}, \bibinfo {author} {\bibfnamefont {C.}~\bibnamefont {Zachreson}},
  \bibinfo {author} {\bibfnamefont {A.~M.}\ \bibnamefont {Berhane}}, \bibinfo
  {author} {\bibfnamefont {K.}~\bibnamefont {Bray}}, \bibinfo {author}
  {\bibfnamefont {R.~G.}\ \bibnamefont {Sandstrom}}, \bibinfo {author}
  {\bibfnamefont {L.~H.}\ \bibnamefont {Li}}, \bibinfo {author} {\bibfnamefont
  {T.}~\bibnamefont {Taniguchi}}, \bibinfo {author} {\bibfnamefont
  {K.}~\bibnamefont {Watanabe}}, \bibinfo {author} {\bibfnamefont
  {I.}~\bibnamefont {Aharonovich}},\ and\ \bibinfo {author} {\bibfnamefont
  {M.}~\bibnamefont {Toth}},\ }\bibfield  {title} {\bibinfo {title} {Quantum
  emission from defects in single-crystalline hexagonal boron nitride},\
  }\href@noop {} {\bibfield  {journal} {\bibinfo  {journal} {Phys. Rev. Appl.}\
  }\textbf {\bibinfo {volume} {5}},\ \bibinfo {pages} {034005} (\bibinfo {year}
  {2016}{\natexlab{b}})}\BibitemShut {NoStop}%
\bibitem [{\citenamefont {Mart{\'\i}nez}\ \emph {et~al.}(2016)\citenamefont
  {Mart{\'\i}nez}, \citenamefont {Pelini}, \citenamefont {Waselowski},
  \citenamefont {Maze}, \citenamefont {Gil}, \citenamefont {Cassabois},\ and\
  \citenamefont {Jacques}}]{zpl_pse_tau_spe_martinez2016efficient}%
  \BibitemOpen
  \bibfield  {author} {\bibinfo {author} {\bibfnamefont {L.}~\bibnamefont
  {Mart{\'\i}nez}}, \bibinfo {author} {\bibfnamefont {T.}~\bibnamefont
  {Pelini}}, \bibinfo {author} {\bibfnamefont {V.}~\bibnamefont {Waselowski}},
  \bibinfo {author} {\bibfnamefont {J.}~\bibnamefont {Maze}}, \bibinfo {author}
  {\bibfnamefont {B.}~\bibnamefont {Gil}}, \bibinfo {author} {\bibfnamefont
  {G.}~\bibnamefont {Cassabois}},\ and\ \bibinfo {author} {\bibfnamefont
  {V.}~\bibnamefont {Jacques}},\ }\bibfield  {title} {\bibinfo {title}
  {Efficient single photon emission from a high-purity hexagonal boron nitride
  crystal},\ }\href@noop {} {\bibfield  {journal} {\bibinfo  {journal} {Phys.
  Rev. B}\ }\textbf {\bibinfo {volume} {94}},\ \bibinfo {pages} {121405}
  (\bibinfo {year} {2016})}\BibitemShut {NoStop}%
\bibitem [{\citenamefont {Dietrich}\ \emph {et~al.}(2018)\citenamefont
  {Dietrich}, \citenamefont {B{\"u}rk}, \citenamefont {Steiger}, \citenamefont
  {Antoniuk}, \citenamefont {Tran}, \citenamefont {Nguyen}, \citenamefont
  {Aharonovich}, \citenamefont {Jelezko},\ and\ \citenamefont
  {Kubanek}}]{zpl_pse_tau_spe_dietrich2018observation}%
  \BibitemOpen
  \bibfield  {author} {\bibinfo {author} {\bibfnamefont {A.}~\bibnamefont
  {Dietrich}}, \bibinfo {author} {\bibfnamefont {M.}~\bibnamefont {B{\"u}rk}},
  \bibinfo {author} {\bibfnamefont {E.~S.}\ \bibnamefont {Steiger}}, \bibinfo
  {author} {\bibfnamefont {L.}~\bibnamefont {Antoniuk}}, \bibinfo {author}
  {\bibfnamefont {T.~T.}\ \bibnamefont {Tran}}, \bibinfo {author}
  {\bibfnamefont {M.}~\bibnamefont {Nguyen}}, \bibinfo {author} {\bibfnamefont
  {I.}~\bibnamefont {Aharonovich}}, \bibinfo {author} {\bibfnamefont
  {F.}~\bibnamefont {Jelezko}},\ and\ \bibinfo {author} {\bibfnamefont
  {A.}~\bibnamefont {Kubanek}},\ }\bibfield  {title} {\bibinfo {title}
  {Observation of fourier transform limited lines in hexagonal boron nitride},\
  }\href@noop {} {\bibfield  {journal} {\bibinfo  {journal} {Phys. Rev. B}\
  }\textbf {\bibinfo {volume} {98}},\ \bibinfo {pages} {081414} (\bibinfo
  {year} {2018})}\BibitemShut {NoStop}%
\bibitem [{\citenamefont {Mendelson}\ \emph {et~al.}(2020)\citenamefont
  {Mendelson}, \citenamefont {Chugh}, \citenamefont {Reimers}, \citenamefont
  {Cheng}, \citenamefont {Gottscholl}, \citenamefont {Long}, \citenamefont
  {Mellor}, \citenamefont {Zettl}, \citenamefont {Dyakonov}, \citenamefont
  {Beton} \emph
  {et~al.}}]{zpl_raman_pse_spe_linpol_spinpol_mendelson2020identifying}%
  \BibitemOpen
  \bibfield  {author} {\bibinfo {author} {\bibfnamefont {N.}~\bibnamefont
  {Mendelson}}, \bibinfo {author} {\bibfnamefont {D.}~\bibnamefont {Chugh}},
  \bibinfo {author} {\bibfnamefont {J.~R.}\ \bibnamefont {Reimers}}, \bibinfo
  {author} {\bibfnamefont {T.~S.}\ \bibnamefont {Cheng}}, \bibinfo {author}
  {\bibfnamefont {A.}~\bibnamefont {Gottscholl}}, \bibinfo {author}
  {\bibfnamefont {H.}~\bibnamefont {Long}}, \bibinfo {author} {\bibfnamefont
  {C.~J.}\ \bibnamefont {Mellor}}, \bibinfo {author} {\bibfnamefont
  {A.}~\bibnamefont {Zettl}}, \bibinfo {author} {\bibfnamefont
  {V.}~\bibnamefont {Dyakonov}}, \bibinfo {author} {\bibfnamefont {P.~H.}\
  \bibnamefont {Beton}}, \emph {et~al.},\ }\bibfield  {title} {\bibinfo {title}
  {Identifying carbon as the source of visible single-photon emission from
  hexagonal boron nitride},\ }\href@noop {} {\bibfield  {journal} {\bibinfo
  {journal} {Nat. Mater.}\ ,\ \bibinfo {pages} {1}} (\bibinfo {year}
  {2020})}\BibitemShut {NoStop}%
\bibitem [{\citenamefont {Hayee}\ \emph {et~al.}(2020)\citenamefont {Hayee},
  \citenamefont {Yu}, \citenamefont {Zhang}, \citenamefont {Ciccarino},
  \citenamefont {Nguyen}, \citenamefont {Marshall}, \citenamefont
  {Aharonovich}, \citenamefont {Vu{\v{c}}kovi{\'c}}, \citenamefont {Narang},
  \citenamefont {Heinz} \emph {et~al.}}]{zpl_pse_spe_hayee2020revealing}%
  \BibitemOpen
  \bibfield  {author} {\bibinfo {author} {\bibfnamefont {F.}~\bibnamefont
  {Hayee}}, \bibinfo {author} {\bibfnamefont {L.}~\bibnamefont {Yu}}, \bibinfo
  {author} {\bibfnamefont {J.~L.}\ \bibnamefont {Zhang}}, \bibinfo {author}
  {\bibfnamefont {C.~J.}\ \bibnamefont {Ciccarino}}, \bibinfo {author}
  {\bibfnamefont {M.}~\bibnamefont {Nguyen}}, \bibinfo {author} {\bibfnamefont
  {A.~F.}\ \bibnamefont {Marshall}}, \bibinfo {author} {\bibfnamefont
  {I.}~\bibnamefont {Aharonovich}}, \bibinfo {author} {\bibfnamefont
  {J.}~\bibnamefont {Vu{\v{c}}kovi{\'c}}}, \bibinfo {author} {\bibfnamefont
  {P.}~\bibnamefont {Narang}}, \bibinfo {author} {\bibfnamefont {T.~F.}\
  \bibnamefont {Heinz}}, \emph {et~al.},\ }\bibfield  {title} {\bibinfo {title}
  {Revealing multiple classes of stable quantum emitters in hexagonal boron
  nitride with correlated optical and electron microscopy},\ }\href@noop {}
  {\bibfield  {journal} {\bibinfo  {journal} {Nat. Mater.}\ }\textbf {\bibinfo
  {volume} {19}},\ \bibinfo {pages} {534} (\bibinfo {year} {2020})}\BibitemShut
  {NoStop}%
\bibitem [{\citenamefont {Nguyen}\ \emph {et~al.}(2018)\citenamefont {Nguyen},
  \citenamefont {Kim}, \citenamefont {Tran}, \citenamefont {Xu}, \citenamefont
  {Kianinia}, \citenamefont {Toth},\ and\ \citenamefont
  {Aharonovich}}]{zpl_pse_dw_tau_qy_spe_linpol_nguyen2018nanoassembly}%
  \BibitemOpen
  \bibfield  {author} {\bibinfo {author} {\bibfnamefont {M.}~\bibnamefont
  {Nguyen}}, \bibinfo {author} {\bibfnamefont {S.}~\bibnamefont {Kim}},
  \bibinfo {author} {\bibfnamefont {T.~T.}\ \bibnamefont {Tran}}, \bibinfo
  {author} {\bibfnamefont {Z.-Q.}\ \bibnamefont {Xu}}, \bibinfo {author}
  {\bibfnamefont {M.}~\bibnamefont {Kianinia}}, \bibinfo {author}
  {\bibfnamefont {M.}~\bibnamefont {Toth}},\ and\ \bibinfo {author}
  {\bibfnamefont {I.}~\bibnamefont {Aharonovich}},\ }\bibfield  {title}
  {\bibinfo {title} {Nanoassembly of quantum emitters in hexagonal boron
  nitride and gold nanospheres},\ }\href@noop {} {\bibfield  {journal}
  {\bibinfo  {journal} {Nanoscale}\ }\textbf {\bibinfo {volume} {10}},\
  \bibinfo {pages} {2267} (\bibinfo {year} {2018})}\BibitemShut {NoStop}%
\bibitem [{\citenamefont {Xu}\ \emph {et~al.}(2018)\citenamefont {Xu},
  \citenamefont {Elbadawi}, \citenamefont {Tran}, \citenamefont {Kianinia},
  \citenamefont {Li}, \citenamefont {Liu}, \citenamefont {Hoffman},
  \citenamefont {Nguyen}, \citenamefont {Kim}, \citenamefont {Edgar} \emph
  {et~al.}}]{zpl_raman_tau_spe_linpol_xu2018single}%
  \BibitemOpen
  \bibfield  {author} {\bibinfo {author} {\bibfnamefont {Z.-Q.}\ \bibnamefont
  {Xu}}, \bibinfo {author} {\bibfnamefont {C.}~\bibnamefont {Elbadawi}},
  \bibinfo {author} {\bibfnamefont {T.~T.}\ \bibnamefont {Tran}}, \bibinfo
  {author} {\bibfnamefont {M.}~\bibnamefont {Kianinia}}, \bibinfo {author}
  {\bibfnamefont {X.}~\bibnamefont {Li}}, \bibinfo {author} {\bibfnamefont
  {D.}~\bibnamefont {Liu}}, \bibinfo {author} {\bibfnamefont {T.~B.}\
  \bibnamefont {Hoffman}}, \bibinfo {author} {\bibfnamefont {M.}~\bibnamefont
  {Nguyen}}, \bibinfo {author} {\bibfnamefont {S.}~\bibnamefont {Kim}},
  \bibinfo {author} {\bibfnamefont {J.~H.}\ \bibnamefont {Edgar}}, \emph
  {et~al.},\ }\bibfield  {title} {\bibinfo {title} {Single photon emission from
  plasma treated 2d hexagonal boron nitride},\ }\href@noop {} {\bibfield
  {journal} {\bibinfo  {journal} {Nanoscale}\ }\textbf {\bibinfo {volume}
  {10}},\ \bibinfo {pages} {7957} (\bibinfo {year} {2018})}\BibitemShut
  {NoStop}%
\bibitem [{\citenamefont {Khatri}\ \emph {et~al.}(2020)\citenamefont {Khatri},
  \citenamefont {Ramsay}, \citenamefont {Malein}, \citenamefont {Chong},\ and\
  \citenamefont {Luxmoore}}]{zpl_pse_tau_spe_linpol_khatri2020optical}%
  \BibitemOpen
  \bibfield  {author} {\bibinfo {author} {\bibfnamefont {P.}~\bibnamefont
  {Khatri}}, \bibinfo {author} {\bibfnamefont {A.~J.}\ \bibnamefont {Ramsay}},
  \bibinfo {author} {\bibfnamefont {R.~N.~E.}\ \bibnamefont {Malein}}, \bibinfo
  {author} {\bibfnamefont {H.~M.}\ \bibnamefont {Chong}},\ and\ \bibinfo
  {author} {\bibfnamefont {I.~J.}\ \bibnamefont {Luxmoore}},\ }\bibfield
  {title} {\bibinfo {title} {Optical gating of photoluminescence from color
  centers in hexagonal boron nitride},\ }\href@noop {} {\bibfield  {journal}
  {\bibinfo  {journal} {Nano Lett.}\ }\textbf {\bibinfo {volume} {20}},\
  \bibinfo {pages} {4256} (\bibinfo {year} {2020})}\BibitemShut {NoStop}%
\bibitem [{\citenamefont {Lazi{\'c}}\ \emph {et~al.}(2019)\citenamefont
  {Lazi{\'c}}, \citenamefont {Espinha}, \citenamefont {Yanguas}, \citenamefont
  {Gibaja}, \citenamefont {Zamora}, \citenamefont {Ares}, \citenamefont
  {Chhowalla}, \citenamefont {Paz}, \citenamefont {Burgos}, \citenamefont
  {Hern{\'a}ndez-M{\'\i}nguez} \emph
  {et~al.}}]{zpl_tau_spe_linpol_lazic2019dynamically}%
  \BibitemOpen
  \bibfield  {author} {\bibinfo {author} {\bibfnamefont {S.}~\bibnamefont
  {Lazi{\'c}}}, \bibinfo {author} {\bibfnamefont {A.}~\bibnamefont {Espinha}},
  \bibinfo {author} {\bibfnamefont {S.~P.}\ \bibnamefont {Yanguas}}, \bibinfo
  {author} {\bibfnamefont {C.}~\bibnamefont {Gibaja}}, \bibinfo {author}
  {\bibfnamefont {F.}~\bibnamefont {Zamora}}, \bibinfo {author} {\bibfnamefont
  {P.}~\bibnamefont {Ares}}, \bibinfo {author} {\bibfnamefont {M.}~\bibnamefont
  {Chhowalla}}, \bibinfo {author} {\bibfnamefont {W.~S.}\ \bibnamefont {Paz}},
  \bibinfo {author} {\bibfnamefont {J.~J.~P.}\ \bibnamefont {Burgos}}, \bibinfo
  {author} {\bibfnamefont {A.}~\bibnamefont {Hern{\'a}ndez-M{\'\i}nguez}},
  \emph {et~al.},\ }\bibfield  {title} {\bibinfo {title} {Dynamically tuned
  ton-classical light emission from atomic defects in hexagonal boron
  nitride},\ }\href@noop {} {\bibfield  {journal} {\bibinfo  {journal} {Commun.
  Phys.}\ }\textbf {\bibinfo {volume} {2}},\ \bibinfo {pages} {1} (\bibinfo
  {year} {2019})}\BibitemShut {NoStop}%
\bibitem [{\citenamefont {Liu}\ \emph {et~al.}(2020)\citenamefont {Liu},
  \citenamefont {Wang}, \citenamefont {Li}, \citenamefont {Yu}, \citenamefont
  {Ke}, \citenamefont {Meng}, \citenamefont {Tang}, \citenamefont {Li},\ and\
  \citenamefont {Guo}}]{zpl_raman_pse_spe_linpol_liu2020ultrastable}%
  \BibitemOpen
  \bibfield  {author} {\bibinfo {author} {\bibfnamefont {W.}~\bibnamefont
  {Liu}}, \bibinfo {author} {\bibfnamefont {Y.-T.}\ \bibnamefont {Wang}},
  \bibinfo {author} {\bibfnamefont {Z.-P.}\ \bibnamefont {Li}}, \bibinfo
  {author} {\bibfnamefont {S.}~\bibnamefont {Yu}}, \bibinfo {author}
  {\bibfnamefont {Z.-J.}\ \bibnamefont {Ke}}, \bibinfo {author} {\bibfnamefont
  {Y.}~\bibnamefont {Meng}}, \bibinfo {author} {\bibfnamefont {J.-S.}\
  \bibnamefont {Tang}}, \bibinfo {author} {\bibfnamefont {C.-F.}\ \bibnamefont
  {Li}},\ and\ \bibinfo {author} {\bibfnamefont {G.-C.}\ \bibnamefont {Guo}},\
  }\bibfield  {title} {\bibinfo {title} {An ultrastable and robust
  single-photon emitter in hexagonal boron nitride},\ }\href@noop {} {\bibfield
   {journal} {\bibinfo  {journal} {Physica E Low Dimens. Syst. Nanostruct.}\
  }\textbf {\bibinfo {volume} {124}},\ \bibinfo {pages} {114251} (\bibinfo
  {year} {2020})}\BibitemShut {NoStop}%
\bibitem [{\citenamefont {Vogl}\ \emph {et~al.}(2018)\citenamefont {Vogl},
  \citenamefont {Campbell}, \citenamefont {Buchler}, \citenamefont {Lu},\ and\
  \citenamefont {Lam}}]{zpl_pse_tau_spe_vogl2018fabrication}%
  \BibitemOpen
  \bibfield  {author} {\bibinfo {author} {\bibfnamefont {T.}~\bibnamefont
  {Vogl}}, \bibinfo {author} {\bibfnamefont {G.}~\bibnamefont {Campbell}},
  \bibinfo {author} {\bibfnamefont {B.~C.}\ \bibnamefont {Buchler}}, \bibinfo
  {author} {\bibfnamefont {Y.}~\bibnamefont {Lu}},\ and\ \bibinfo {author}
  {\bibfnamefont {P.~K.}\ \bibnamefont {Lam}},\ }\bibfield  {title} {\bibinfo
  {title} {Fabrication and deterministic transfer of high-quality quantum
  emitters in hexagonal boron nitride},\ }\href@noop {} {\bibfield  {journal}
  {\bibinfo  {journal} {ACS Photonics}\ }\textbf {\bibinfo {volume} {5}},\
  \bibinfo {pages} {2305} (\bibinfo {year} {2018})}\BibitemShut {NoStop}%
\bibitem [{\citenamefont {Tran}\ \emph {et~al.}(2018)\citenamefont {Tran},
  \citenamefont {Kianinia}, \citenamefont {Nguyen}, \citenamefont {Kim},
  \citenamefont {Xu}, \citenamefont {Kubanek}, \citenamefont {Toth},\ and\
  \citenamefont {Aharonovich}}]{zpl_pse_tau_spe_tran2018resonant}%
  \BibitemOpen
  \bibfield  {author} {\bibinfo {author} {\bibfnamefont {T.~T.}\ \bibnamefont
  {Tran}}, \bibinfo {author} {\bibfnamefont {M.}~\bibnamefont {Kianinia}},
  \bibinfo {author} {\bibfnamefont {M.}~\bibnamefont {Nguyen}}, \bibinfo
  {author} {\bibfnamefont {S.}~\bibnamefont {Kim}}, \bibinfo {author}
  {\bibfnamefont {Z.-Q.}\ \bibnamefont {Xu}}, \bibinfo {author} {\bibfnamefont
  {A.}~\bibnamefont {Kubanek}}, \bibinfo {author} {\bibfnamefont
  {M.}~\bibnamefont {Toth}},\ and\ \bibinfo {author} {\bibfnamefont
  {I.}~\bibnamefont {Aharonovich}},\ }\bibfield  {title} {\bibinfo {title}
  {Resonant excitation of quantum emitters in hexagonal boron nitride},\
  }\href@noop {} {\bibfield  {journal} {\bibinfo  {journal} {ACS Photonics}\
  }\textbf {\bibinfo {volume} {5}},\ \bibinfo {pages} {295} (\bibinfo {year}
  {2018})}\BibitemShut {NoStop}%
\bibitem [{\citenamefont {Mendelson}\ \emph {et~al.}(2019)\citenamefont
  {Mendelson}, \citenamefont {Xu}, \citenamefont {Tran}, \citenamefont
  {Kianinia}, \citenamefont {Scott}, \citenamefont {Bradac}, \citenamefont
  {Aharonovich},\ and\ \citenamefont
  {Toth}}]{zpl_raman_pse_spe_mendelson2019engineering}%
  \BibitemOpen
  \bibfield  {author} {\bibinfo {author} {\bibfnamefont {N.}~\bibnamefont
  {Mendelson}}, \bibinfo {author} {\bibfnamefont {Z.-Q.}\ \bibnamefont {Xu}},
  \bibinfo {author} {\bibfnamefont {T.~T.}\ \bibnamefont {Tran}}, \bibinfo
  {author} {\bibfnamefont {M.}~\bibnamefont {Kianinia}}, \bibinfo {author}
  {\bibfnamefont {J.}~\bibnamefont {Scott}}, \bibinfo {author} {\bibfnamefont
  {C.}~\bibnamefont {Bradac}}, \bibinfo {author} {\bibfnamefont
  {I.}~\bibnamefont {Aharonovich}},\ and\ \bibinfo {author} {\bibfnamefont
  {M.}~\bibnamefont {Toth}},\ }\bibfield  {title} {\bibinfo {title}
  {Engineering and tuning of quantum emitters in few-layer hexagonal boron
  nitride},\ }\href@noop {} {\bibfield  {journal} {\bibinfo  {journal} {ACS
  Nano}\ }\textbf {\bibinfo {volume} {13}},\ \bibinfo {pages} {3132} (\bibinfo
  {year} {2019})}\BibitemShut {NoStop}%
\bibitem [{\citenamefont {Exarhos}\ \emph {et~al.}(2017)\citenamefont
  {Exarhos}, \citenamefont {Hopper}, \citenamefont {Grote}, \citenamefont
  {Alkauskas},\ and\ \citenamefont
  {Bassett}}]{zpl_pse_hr_tau_spe_linpol_exarhos2017optical}%
  \BibitemOpen
  \bibfield  {author} {\bibinfo {author} {\bibfnamefont {A.~L.}\ \bibnamefont
  {Exarhos}}, \bibinfo {author} {\bibfnamefont {D.~A.}\ \bibnamefont {Hopper}},
  \bibinfo {author} {\bibfnamefont {R.~R.}\ \bibnamefont {Grote}}, \bibinfo
  {author} {\bibfnamefont {A.}~\bibnamefont {Alkauskas}},\ and\ \bibinfo
  {author} {\bibfnamefont {L.~C.}\ \bibnamefont {Bassett}},\ }\bibfield
  {title} {\bibinfo {title} {Optical signatures of quantum emitters in
  suspended hexagonal boron nitride},\ }\href@noop {} {\bibfield  {journal}
  {\bibinfo  {journal} {ACS Nano}\ }\textbf {\bibinfo {volume} {11}},\ \bibinfo
  {pages} {3328} (\bibinfo {year} {2017})}\BibitemShut {NoStop}%
\bibitem [{\citenamefont {Tran}\ \emph
  {et~al.}(2016{\natexlab{c}})\citenamefont {Tran}, \citenamefont {Elbadawi},
  \citenamefont {Totonjian}, \citenamefont {Lobo}, \citenamefont {Grosso},
  \citenamefont {Moon}, \citenamefont {Englund}, \citenamefont {Ford},
  \citenamefont {Aharonovich},\ and\ \citenamefont
  {Toth}}]{zpl_pse_tau_spe_tran2016robust}%
  \BibitemOpen
  \bibfield  {author} {\bibinfo {author} {\bibfnamefont {T.~T.}\ \bibnamefont
  {Tran}}, \bibinfo {author} {\bibfnamefont {C.}~\bibnamefont {Elbadawi}},
  \bibinfo {author} {\bibfnamefont {D.}~\bibnamefont {Totonjian}}, \bibinfo
  {author} {\bibfnamefont {C.~J.}\ \bibnamefont {Lobo}}, \bibinfo {author}
  {\bibfnamefont {G.}~\bibnamefont {Grosso}}, \bibinfo {author} {\bibfnamefont
  {H.}~\bibnamefont {Moon}}, \bibinfo {author} {\bibfnamefont {D.~R.}\
  \bibnamefont {Englund}}, \bibinfo {author} {\bibfnamefont {M.~J.}\
  \bibnamefont {Ford}}, \bibinfo {author} {\bibfnamefont {I.}~\bibnamefont
  {Aharonovich}},\ and\ \bibinfo {author} {\bibfnamefont {M.}~\bibnamefont
  {Toth}},\ }\bibfield  {title} {\bibinfo {title} {Robust multicolor single
  photon emission from point defects in hexagonal boron nitride},\ }\href@noop
  {} {\bibfield  {journal} {\bibinfo  {journal} {ACS Nano}\ }\textbf {\bibinfo
  {volume} {10}},\ \bibinfo {pages} {7331} (\bibinfo {year}
  {2016}{\natexlab{c}})}\BibitemShut {NoStop}%
\bibitem [{\citenamefont {Ngoc My~Duong}\ \emph {et~al.}(2018)\citenamefont
  {Ngoc My~Duong}, \citenamefont {Nguyen}, \citenamefont {Kianinia},
  \citenamefont {Ohshima}, \citenamefont {Abe}, \citenamefont {Watanabe},
  \citenamefont {Taniguchi}, \citenamefont {Edgar}, \citenamefont
  {Aharonovich},\ and\ \citenamefont {Toth}}]{zpl_pse_spe_ngoc2018effects}%
  \BibitemOpen
  \bibfield  {author} {\bibinfo {author} {\bibfnamefont {H.}~\bibnamefont {Ngoc
  My~Duong}}, \bibinfo {author} {\bibfnamefont {M.~A.~P.}\ \bibnamefont
  {Nguyen}}, \bibinfo {author} {\bibfnamefont {M.}~\bibnamefont {Kianinia}},
  \bibinfo {author} {\bibfnamefont {T.}~\bibnamefont {Ohshima}}, \bibinfo
  {author} {\bibfnamefont {H.}~\bibnamefont {Abe}}, \bibinfo {author}
  {\bibfnamefont {K.}~\bibnamefont {Watanabe}}, \bibinfo {author}
  {\bibfnamefont {T.}~\bibnamefont {Taniguchi}}, \bibinfo {author}
  {\bibfnamefont {J.~H.}\ \bibnamefont {Edgar}}, \bibinfo {author}
  {\bibfnamefont {I.}~\bibnamefont {Aharonovich}},\ and\ \bibinfo {author}
  {\bibfnamefont {M.}~\bibnamefont {Toth}},\ }\bibfield  {title} {\bibinfo
  {title} {Effects of high-energy electron irradiation on quantum emitters in
  hexagonal boron nitride},\ }\href@noop {} {\bibfield  {journal} {\bibinfo
  {journal} {ACS Appl. Mater. Interfaces}\ }\textbf {\bibinfo {volume} {10}},\
  \bibinfo {pages} {24886} (\bibinfo {year} {2018})}\BibitemShut {NoStop}%
\bibitem [{\citenamefont {Choi}\ \emph {et~al.}(2016)\citenamefont {Choi},
  \citenamefont {Tran}, \citenamefont {Elbadawi}, \citenamefont {Lobo},
  \citenamefont {Wang}, \citenamefont {Juodkazis}, \citenamefont {Seniutinas},
  \citenamefont {Toth},\ and\ \citenamefont
  {Aharonovich}}]{zpl_pse_spe_choi2016engineering}%
  \BibitemOpen
  \bibfield  {author} {\bibinfo {author} {\bibfnamefont {S.}~\bibnamefont
  {Choi}}, \bibinfo {author} {\bibfnamefont {T.~T.}\ \bibnamefont {Tran}},
  \bibinfo {author} {\bibfnamefont {C.}~\bibnamefont {Elbadawi}}, \bibinfo
  {author} {\bibfnamefont {C.}~\bibnamefont {Lobo}}, \bibinfo {author}
  {\bibfnamefont {X.}~\bibnamefont {Wang}}, \bibinfo {author} {\bibfnamefont
  {S.}~\bibnamefont {Juodkazis}}, \bibinfo {author} {\bibfnamefont
  {G.}~\bibnamefont {Seniutinas}}, \bibinfo {author} {\bibfnamefont
  {M.}~\bibnamefont {Toth}},\ and\ \bibinfo {author} {\bibfnamefont
  {I.}~\bibnamefont {Aharonovich}},\ }\bibfield  {title} {\bibinfo {title}
  {Engineering and localization of quantum emitters in large hexagonal boron
  nitride layers},\ }\href@noop {} {\bibfield  {journal} {\bibinfo  {journal}
  {ACS Appl. Mater. Interfaces}\ }\textbf {\bibinfo {volume} {8}},\ \bibinfo
  {pages} {29642} (\bibinfo {year} {2016})}\BibitemShut {NoStop}%
\bibitem [{\citenamefont {Grosso}\ \emph {et~al.}(2020)\citenamefont {Grosso},
  \citenamefont {Moon}, \citenamefont {Ciccarino}, \citenamefont {Flick},
  \citenamefont {Mendelson}, \citenamefont {Mennel}, \citenamefont {Toth},
  \citenamefont {Aharonovich}, \citenamefont {Narang},\ and\ \citenamefont
  {Englund}}]{zpl_tau_spe_grosso2020low}%
  \BibitemOpen
  \bibfield  {author} {\bibinfo {author} {\bibfnamefont {G.}~\bibnamefont
  {Grosso}}, \bibinfo {author} {\bibfnamefont {H.}~\bibnamefont {Moon}},
  \bibinfo {author} {\bibfnamefont {C.~J.}\ \bibnamefont {Ciccarino}}, \bibinfo
  {author} {\bibfnamefont {J.}~\bibnamefont {Flick}}, \bibinfo {author}
  {\bibfnamefont {N.}~\bibnamefont {Mendelson}}, \bibinfo {author}
  {\bibfnamefont {L.}~\bibnamefont {Mennel}}, \bibinfo {author} {\bibfnamefont
  {M.}~\bibnamefont {Toth}}, \bibinfo {author} {\bibfnamefont {I.}~\bibnamefont
  {Aharonovich}}, \bibinfo {author} {\bibfnamefont {P.}~\bibnamefont
  {Narang}},\ and\ \bibinfo {author} {\bibfnamefont {D.~R.}\ \bibnamefont
  {Englund}},\ }\bibfield  {title} {\bibinfo {title} {Low-temperature
  electron--phonon interaction of quantum emitters in hexagonal boron
  nitride},\ }\href@noop {} {\bibfield  {journal} {\bibinfo  {journal} {ACS
  Photonics}\ }\textbf {\bibinfo {volume} {7}},\ \bibinfo {pages} {1410}
  (\bibinfo {year} {2020})}\BibitemShut {NoStop}%
\bibitem [{\citenamefont {Neu}\ \emph {et~al.}(2011)\citenamefont {Neu},
  \citenamefont {Steinmetz}, \citenamefont {Riedrich-M{\"o}ller}, \citenamefont
  {Gsell}, \citenamefont {Fischer}, \citenamefont {Schreck},\ and\
  \citenamefont {Becher}}]{neu2011single}%
  \BibitemOpen
  \bibfield  {author} {\bibinfo {author} {\bibfnamefont {E.}~\bibnamefont
  {Neu}}, \bibinfo {author} {\bibfnamefont {D.}~\bibnamefont {Steinmetz}},
  \bibinfo {author} {\bibfnamefont {J.}~\bibnamefont {Riedrich-M{\"o}ller}},
  \bibinfo {author} {\bibfnamefont {S.}~\bibnamefont {Gsell}}, \bibinfo
  {author} {\bibfnamefont {M.}~\bibnamefont {Fischer}}, \bibinfo {author}
  {\bibfnamefont {M.}~\bibnamefont {Schreck}},\ and\ \bibinfo {author}
  {\bibfnamefont {C.}~\bibnamefont {Becher}},\ }\bibfield  {title} {\bibinfo
  {title} {Single photon emission from silicon-vacancy colour centres in
  chemical vapour deposition nano-diamonds on iridium},\ }\href@noop {}
  {\bibfield  {journal} {\bibinfo  {journal} {New J. Phys.}\ }\textbf {\bibinfo
  {volume} {13}},\ \bibinfo {pages} {025012} (\bibinfo {year}
  {2011})}\BibitemShut {NoStop}%
\bibitem [{\citenamefont {Konthasinghe}\ \emph {et~al.}(2019)\citenamefont
  {Konthasinghe}, \citenamefont {Chakraborty}, \citenamefont {Mathur},
  \citenamefont {Qiu}, \citenamefont {Mukherjee}, \citenamefont {Fuchs},\ and\
  \citenamefont {Vamivakas}}]{zpl_pse_linpol_konthasinghe2019rabi}%
  \BibitemOpen
  \bibfield  {author} {\bibinfo {author} {\bibfnamefont {K.}~\bibnamefont
  {Konthasinghe}}, \bibinfo {author} {\bibfnamefont {C.}~\bibnamefont
  {Chakraborty}}, \bibinfo {author} {\bibfnamefont {N.}~\bibnamefont {Mathur}},
  \bibinfo {author} {\bibfnamefont {L.}~\bibnamefont {Qiu}}, \bibinfo {author}
  {\bibfnamefont {A.}~\bibnamefont {Mukherjee}}, \bibinfo {author}
  {\bibfnamefont {G.~D.}\ \bibnamefont {Fuchs}},\ and\ \bibinfo {author}
  {\bibfnamefont {A.~N.}\ \bibnamefont {Vamivakas}},\ }\bibfield  {title}
  {\bibinfo {title} {Rabi oscillations and resonance fluorescence from a single
  hexagonal boron nitride quantum emitter},\ }\href@noop {} {\bibfield
  {journal} {\bibinfo  {journal} {Optica}\ }\textbf {\bibinfo {volume} {6}},\
  \bibinfo {pages} {542} (\bibinfo {year} {2019})}\BibitemShut {NoStop}%
\bibitem [{\citenamefont {Wang}\ \emph {et~al.}(2018)\citenamefont {Wang},
  \citenamefont {Zhang}, \citenamefont {Zhao}, \citenamefont {Luo},
  \citenamefont {Wong}, \citenamefont {Wang}, \citenamefont {Wan},
  \citenamefont {Venkatesan}, \citenamefont {Pennycook}, \citenamefont {Loh}
  \emph {et~al.}}]{zpl_raman_pse_hr_dw_wang2018photoluminescence}%
  \BibitemOpen
  \bibfield  {author} {\bibinfo {author} {\bibfnamefont {Q.}~\bibnamefont
  {Wang}}, \bibinfo {author} {\bibfnamefont {Q.}~\bibnamefont {Zhang}},
  \bibinfo {author} {\bibfnamefont {X.}~\bibnamefont {Zhao}}, \bibinfo {author}
  {\bibfnamefont {X.}~\bibnamefont {Luo}}, \bibinfo {author} {\bibfnamefont
  {C.~P.~Y.}\ \bibnamefont {Wong}}, \bibinfo {author} {\bibfnamefont
  {J.}~\bibnamefont {Wang}}, \bibinfo {author} {\bibfnamefont {D.}~\bibnamefont
  {Wan}}, \bibinfo {author} {\bibfnamefont {T.}~\bibnamefont {Venkatesan}},
  \bibinfo {author} {\bibfnamefont {S.~J.}\ \bibnamefont {Pennycook}}, \bibinfo
  {author} {\bibfnamefont {K.~P.}\ \bibnamefont {Loh}}, \emph {et~al.},\
  }\bibfield  {title} {\bibinfo {title} {Photoluminescence upconversion by
  defects in hexagonal boron nitride},\ }\href@noop {} {\bibfield  {journal}
  {\bibinfo  {journal} {Nano Lett.}\ }\textbf {\bibinfo {volume} {18}},\
  \bibinfo {pages} {6898} (\bibinfo {year} {2018})}\BibitemShut {NoStop}%
\bibitem [{\citenamefont {Schell}\ \emph {et~al.}(2017)\citenamefont {Schell},
  \citenamefont {Takashima}, \citenamefont {Tran}, \citenamefont
  {Aharonovich},\ and\ \citenamefont
  {Takeuchi}}]{zpl_tau_linpol_schell2017coupling}%
  \BibitemOpen
  \bibfield  {author} {\bibinfo {author} {\bibfnamefont {A.~W.}\ \bibnamefont
  {Schell}}, \bibinfo {author} {\bibfnamefont {H.}~\bibnamefont {Takashima}},
  \bibinfo {author} {\bibfnamefont {T.~T.}\ \bibnamefont {Tran}}, \bibinfo
  {author} {\bibfnamefont {I.}~\bibnamefont {Aharonovich}},\ and\ \bibinfo
  {author} {\bibfnamefont {S.}~\bibnamefont {Takeuchi}},\ }\bibfield  {title}
  {\bibinfo {title} {Coupling quantum emitters in 2d materials with tapered
  fibers},\ }\href@noop {} {\bibfield  {journal} {\bibinfo  {journal} {ACS
  Photonics}\ }\textbf {\bibinfo {volume} {4}},\ \bibinfo {pages} {761}
  (\bibinfo {year} {2017})}\BibitemShut {NoStop}%
\bibitem [{\citenamefont {Exarhos}\ \emph {et~al.}(2019)\citenamefont
  {Exarhos}, \citenamefont {Hopper}, \citenamefont {Patel}, \citenamefont
  {Doherty},\ and\ \citenamefont
  {Bassett}}]{zpl_spe_linpol_spinpol_exarhos2019magnetic}%
  \BibitemOpen
  \bibfield  {author} {\bibinfo {author} {\bibfnamefont {A.~L.}\ \bibnamefont
  {Exarhos}}, \bibinfo {author} {\bibfnamefont {D.~A.}\ \bibnamefont {Hopper}},
  \bibinfo {author} {\bibfnamefont {R.~N.}\ \bibnamefont {Patel}}, \bibinfo
  {author} {\bibfnamefont {M.~W.}\ \bibnamefont {Doherty}},\ and\ \bibinfo
  {author} {\bibfnamefont {L.~C.}\ \bibnamefont {Bassett}},\ }\bibfield
  {title} {\bibinfo {title} {Magnetic-field-dependent quantum emission in
  hexagonal boron nitride at room temperature},\ }\href@noop {} {\bibfield
  {journal} {\bibinfo  {journal} {Nat. Commun.}\ }\textbf {\bibinfo {volume}
  {10}},\ \bibinfo {pages} {1} (\bibinfo {year} {2019})}\BibitemShut {NoStop}%
\bibitem [{\citenamefont {Wu}\ \emph {et~al.}(2019{\natexlab{a}})\citenamefont
  {Wu}, \citenamefont {Rocca},\ and\ \citenamefont
  {Ping}}]{wu2019dimensionality}%
  \BibitemOpen
  \bibfield  {author} {\bibinfo {author} {\bibfnamefont {F.}~\bibnamefont
  {Wu}}, \bibinfo {author} {\bibfnamefont {D.}~\bibnamefont {Rocca}},\ and\
  \bibinfo {author} {\bibfnamefont {Y.}~\bibnamefont {Ping}},\ }\bibfield
  {title} {\bibinfo {title} {Dimensionality and anisotropicity dependence of
  radiative recombination in nanostructured phosphorene},\ }\href@noop {}
  {\bibfield  {journal} {\bibinfo  {journal} {J. Mater. Chem. C}\ }\textbf
  {\bibinfo {volume} {7}},\ \bibinfo {pages} {12891} (\bibinfo {year}
  {2019}{\natexlab{a}})}\BibitemShut {NoStop}%
\bibitem [{\citenamefont {Yim}\ \emph {et~al.}(2020)\citenamefont {Yim},
  \citenamefont {Yu}, \citenamefont {Noh}, \citenamefont {Lee},\ and\
  \citenamefont {Seo}}]{yim2020polarization}%
  \BibitemOpen
  \bibfield  {author} {\bibinfo {author} {\bibfnamefont {D.}~\bibnamefont
  {Yim}}, \bibinfo {author} {\bibfnamefont {M.}~\bibnamefont {Yu}}, \bibinfo
  {author} {\bibfnamefont {G.}~\bibnamefont {Noh}}, \bibinfo {author}
  {\bibfnamefont {J.}~\bibnamefont {Lee}},\ and\ \bibinfo {author}
  {\bibfnamefont {H.}~\bibnamefont {Seo}},\ }\bibfield  {title} {\bibinfo
  {title} {Polarization and localization of single-photon emitters in hexagonal
  boron nitride wrinkles},\ }\href@noop {} {\bibfield  {journal} {\bibinfo
  {journal} {ACS Appl. Mater. Interfaces}\ }\textbf {\bibinfo {volume} {12}},\
  \bibinfo {pages} {36362} (\bibinfo {year} {2020})}\BibitemShut {NoStop}%
\bibitem [{\citenamefont {Krivanek}\ \emph {et~al.}(2010)\citenamefont
  {Krivanek}, \citenamefont {Chisholm}, \citenamefont {Nicolosi}, \citenamefont
  {Pennycook}, \citenamefont {Corbin}, \citenamefont {Dellby}, \citenamefont
  {Murfitt}, \citenamefont {Own}, \citenamefont {Szilagyi}, \citenamefont
  {Oxley} \emph {et~al.}}]{krivanek2010atom}%
  \BibitemOpen
  \bibfield  {author} {\bibinfo {author} {\bibfnamefont {O.~L.}\ \bibnamefont
  {Krivanek}}, \bibinfo {author} {\bibfnamefont {M.~F.}\ \bibnamefont
  {Chisholm}}, \bibinfo {author} {\bibfnamefont {V.}~\bibnamefont {Nicolosi}},
  \bibinfo {author} {\bibfnamefont {T.~J.}\ \bibnamefont {Pennycook}}, \bibinfo
  {author} {\bibfnamefont {G.~J.}\ \bibnamefont {Corbin}}, \bibinfo {author}
  {\bibfnamefont {N.}~\bibnamefont {Dellby}}, \bibinfo {author} {\bibfnamefont
  {M.~F.}\ \bibnamefont {Murfitt}}, \bibinfo {author} {\bibfnamefont {C.~S.}\
  \bibnamefont {Own}}, \bibinfo {author} {\bibfnamefont {Z.~S.}\ \bibnamefont
  {Szilagyi}}, \bibinfo {author} {\bibfnamefont {M.~P.}\ \bibnamefont {Oxley}},
  \emph {et~al.},\ }\bibfield  {title} {\bibinfo {title} {Atom-by-atom
  structural and chemical analysis by annular dark-field electron microscopy},\
  }\href@noop {} {\bibfield  {journal} {\bibinfo  {journal} {Nature}\ }\textbf
  {\bibinfo {volume} {464}},\ \bibinfo {pages} {571} (\bibinfo {year}
  {2010})}\BibitemShut {NoStop}%
\bibitem [{\citenamefont {Sajid}\ and\ \citenamefont
  {Thygesen}(2020)}]{sajid2020vncb}%
  \BibitemOpen
  \bibfield  {author} {\bibinfo {author} {\bibfnamefont {A.}~\bibnamefont
  {Sajid}}\ and\ \bibinfo {author} {\bibfnamefont {K.~S.}\ \bibnamefont
  {Thygesen}},\ }\bibfield  {title} {\bibinfo {title} {Vncb defect as source of
  single photon emission from hexagonal boron nitride},\ }\href@noop {}
  {\bibfield  {journal} {\bibinfo  {journal} {2D Mater.}\ }\textbf {\bibinfo
  {volume} {7}},\ \bibinfo {pages} {031007} (\bibinfo {year}
  {2020})}\BibitemShut {NoStop}%
\bibitem [{\citenamefont {Turiansky}\ and\ \citenamefont {Van~de
  Walle}(2021)}]{turiansky2021impact}%
  \BibitemOpen
  \bibfield  {author} {\bibinfo {author} {\bibfnamefont {M.~E.}\ \bibnamefont
  {Turiansky}}\ and\ \bibinfo {author} {\bibfnamefont {C.~G.}\ \bibnamefont
  {Van~de Walle}},\ }\bibfield  {title} {\bibinfo {title} {Impact of dangling
  bonds on properties of h-bn},\ }\href@noop {} {\bibfield  {journal} {\bibinfo
   {journal} {2D Mater.}\ }\textbf {\bibinfo {volume} {8}},\ \bibinfo {pages}
  {024002} (\bibinfo {year} {2021})}\BibitemShut {NoStop}%
\bibitem [{\citenamefont {Wu}\ \emph {et~al.}(2019{\natexlab{b}})\citenamefont
  {Wu}, \citenamefont {Smart}, \citenamefont {Xu},\ and\ \citenamefont
  {Ping}}]{wu2019carrier}%
  \BibitemOpen
  \bibfield  {author} {\bibinfo {author} {\bibfnamefont {F.}~\bibnamefont
  {Wu}}, \bibinfo {author} {\bibfnamefont {T.~J.}\ \bibnamefont {Smart}},
  \bibinfo {author} {\bibfnamefont {J.}~\bibnamefont {Xu}},\ and\ \bibinfo
  {author} {\bibfnamefont {Y.}~\bibnamefont {Ping}},\ }\bibfield  {title}
  {\bibinfo {title} {Carrier recombination mechanism at defects in wide band
  gap two-dimensional materials from first principles},\ }\href@noop {}
  {\bibfield  {journal} {\bibinfo  {journal} {Phys. Rev. B}\ }\textbf {\bibinfo
  {volume} {100}},\ \bibinfo {pages} {081407} (\bibinfo {year}
  {2019}{\natexlab{b}})}\BibitemShut {NoStop}%
\bibitem [{\citenamefont {Jara}\ \emph {et~al.}(2021)\citenamefont {Jara},
  \citenamefont {Rauch}, \citenamefont {Botti}, \citenamefont {Marques},
  \citenamefont {Norambuena}, \citenamefont {Coto}, \citenamefont
  {Castellanos-Águila}, \citenamefont {Maze},\ and\ \citenamefont
  {Munoz}}]{jara2021first}%
  \BibitemOpen
  \bibfield  {author} {\bibinfo {author} {\bibfnamefont {C.}~\bibnamefont
  {Jara}}, \bibinfo {author} {\bibfnamefont {T.}~\bibnamefont {Rauch}},
  \bibinfo {author} {\bibfnamefont {S.}~\bibnamefont {Botti}}, \bibinfo
  {author} {\bibfnamefont {M.~A.}\ \bibnamefont {Marques}}, \bibinfo {author}
  {\bibfnamefont {A.}~\bibnamefont {Norambuena}}, \bibinfo {author}
  {\bibfnamefont {R.}~\bibnamefont {Coto}}, \bibinfo {author} {\bibfnamefont
  {J.}~\bibnamefont {Castellanos-Águila}}, \bibinfo {author} {\bibfnamefont
  {J.~R.}\ \bibnamefont {Maze}},\ and\ \bibinfo {author} {\bibfnamefont
  {F.}~\bibnamefont {Munoz}},\ }\bibfield  {title} {\bibinfo {title}
  {First-principles identification of single photon emitters based on carbon
  clusters in hexagonal boron nitride},\ }\href@noop {} {\bibfield  {journal}
  {\bibinfo  {journal} {J. Phys. Chem. A}\ }\textbf {\bibinfo {volume} {125}},\
  \bibinfo {pages} {1325} (\bibinfo {year} {2021})}\BibitemShut {NoStop}%
\bibitem [{\citenamefont {Cheng}\ \emph {et~al.}(2017)\citenamefont {Cheng},
  \citenamefont {Zhang}, \citenamefont {Yan}, \citenamefont {Huang},
  \citenamefont {Huang}, \citenamefont {Song}, \citenamefont {Chen},\ and\
  \citenamefont {Tang}}]{cheng2017paramagnetic}%
  \BibitemOpen
  \bibfield  {author} {\bibinfo {author} {\bibfnamefont {G.}~\bibnamefont
  {Cheng}}, \bibinfo {author} {\bibfnamefont {Y.}~\bibnamefont {Zhang}},
  \bibinfo {author} {\bibfnamefont {L.}~\bibnamefont {Yan}}, \bibinfo {author}
  {\bibfnamefont {H.}~\bibnamefont {Huang}}, \bibinfo {author} {\bibfnamefont
  {Q.}~\bibnamefont {Huang}}, \bibinfo {author} {\bibfnamefont
  {Y.}~\bibnamefont {Song}}, \bibinfo {author} {\bibfnamefont {Y.}~\bibnamefont
  {Chen}},\ and\ \bibinfo {author} {\bibfnamefont {Z.}~\bibnamefont {Tang}},\
  }\bibfield  {title} {\bibinfo {title} {A paramagnetic neutral cbvn center in
  hexagonal boron nitride monolayer for spin qubit application},\ }\href@noop
  {} {\bibfield  {journal} {\bibinfo  {journal} {Comput. Mater. Sci.}\ }\textbf
  {\bibinfo {volume} {129}},\ \bibinfo {pages} {247} (\bibinfo {year}
  {2017})}\BibitemShut {NoStop}%
\bibitem [{\citenamefont {Giannozzi}\ \emph {et~al.}(2009)\citenamefont
  {Giannozzi}, \citenamefont {Baroni}, \citenamefont {Bonini}, \citenamefont
  {Calandra}, \citenamefont {Car}, \citenamefont {Cavazzoni}, \citenamefont
  {Ceresoli}, \citenamefont {Chiarotti}, \citenamefont {Cococcioni},
  \citenamefont {Dabo}, \citenamefont {{Dal Corso}}, \citenamefont
  {de~Gironcoli}, \citenamefont {Fabris}, \citenamefont {Fratesi},
  \citenamefont {Gebauer}, \citenamefont {Gerstmann}, \citenamefont
  {Gougoussis}, \citenamefont {Kokalj}, \citenamefont {Lazzeri}, \citenamefont
  {Martin-Samos}, \citenamefont {Marzari}, \citenamefont {Mauri}, \citenamefont
  {Mazzarello}, \citenamefont {Paolini}, \citenamefont {Pasquarello},
  \citenamefont {Paulatto}, \citenamefont {Sbraccia}, \citenamefont {Scandolo},
  \citenamefont {Sclauzero}, \citenamefont {Seitsonen}, \citenamefont
  {Smogunov}, \citenamefont {Umari},\ and\ \citenamefont {Wentzcovitch}}]{QE}%
  \BibitemOpen
  \bibfield  {author} {\bibinfo {author} {\bibfnamefont {P.}~\bibnamefont
  {Giannozzi}}, \bibinfo {author} {\bibfnamefont {S.}~\bibnamefont {Baroni}},
  \bibinfo {author} {\bibfnamefont {N.}~\bibnamefont {Bonini}}, \bibinfo
  {author} {\bibfnamefont {M.}~\bibnamefont {Calandra}}, \bibinfo {author}
  {\bibfnamefont {R.}~\bibnamefont {Car}}, \bibinfo {author} {\bibfnamefont
  {C.}~\bibnamefont {Cavazzoni}}, \bibinfo {author} {\bibfnamefont
  {D.}~\bibnamefont {Ceresoli}}, \bibinfo {author} {\bibfnamefont {G.~L.}\
  \bibnamefont {Chiarotti}}, \bibinfo {author} {\bibfnamefont {M.}~\bibnamefont
  {Cococcioni}}, \bibinfo {author} {\bibfnamefont {I.}~\bibnamefont {Dabo}},
  \bibinfo {author} {\bibfnamefont {A.}~\bibnamefont {{Dal Corso}}}, \bibinfo
  {author} {\bibfnamefont {S.}~\bibnamefont {de~Gironcoli}}, \bibinfo {author}
  {\bibfnamefont {S.}~\bibnamefont {Fabris}}, \bibinfo {author} {\bibfnamefont
  {G.}~\bibnamefont {Fratesi}}, \bibinfo {author} {\bibfnamefont
  {R.}~\bibnamefont {Gebauer}}, \bibinfo {author} {\bibfnamefont
  {U.}~\bibnamefont {Gerstmann}}, \bibinfo {author} {\bibfnamefont
  {C.}~\bibnamefont {Gougoussis}}, \bibinfo {author} {\bibfnamefont
  {A.}~\bibnamefont {Kokalj}}, \bibinfo {author} {\bibfnamefont
  {M.}~\bibnamefont {Lazzeri}}, \bibinfo {author} {\bibfnamefont
  {L.}~\bibnamefont {Martin-Samos}}, \bibinfo {author} {\bibfnamefont
  {N.}~\bibnamefont {Marzari}}, \bibinfo {author} {\bibfnamefont
  {F.}~\bibnamefont {Mauri}}, \bibinfo {author} {\bibfnamefont
  {R.}~\bibnamefont {Mazzarello}}, \bibinfo {author} {\bibfnamefont
  {S.}~\bibnamefont {Paolini}}, \bibinfo {author} {\bibfnamefont
  {A.}~\bibnamefont {Pasquarello}}, \bibinfo {author} {\bibfnamefont
  {L.}~\bibnamefont {Paulatto}}, \bibinfo {author} {\bibfnamefont
  {C.}~\bibnamefont {Sbraccia}}, \bibinfo {author} {\bibfnamefont
  {S.}~\bibnamefont {Scandolo}}, \bibinfo {author} {\bibfnamefont
  {G.}~\bibnamefont {Sclauzero}}, \bibinfo {author} {\bibfnamefont {A.~P.}\
  \bibnamefont {Seitsonen}}, \bibinfo {author} {\bibfnamefont {A.}~\bibnamefont
  {Smogunov}}, \bibinfo {author} {\bibfnamefont {P.}~\bibnamefont {Umari}},\
  and\ \bibinfo {author} {\bibfnamefont {R.~M.}\ \bibnamefont {Wentzcovitch}},\
  }\bibfield  {title} {\bibinfo {title} {{QUANTUM ESPRESSO: A Modular and
  Open-Source Software Project for Quantum Simulations of Materials.}},\
  }\href@noop {} {\bibfield  {journal} {\bibinfo  {journal} {J. Phys.: Condens.
  Matter}\ }\textbf {\bibinfo {volume} {21}},\ \bibinfo {pages} {395502}
  (\bibinfo {year} {2009})}\BibitemShut {NoStop}%
\bibitem [{\citenamefont {Hamann}(2013)}]{ONCV1}%
  \BibitemOpen
  \bibfield  {author} {\bibinfo {author} {\bibfnamefont {D.~R.}\ \bibnamefont
  {Hamann}},\ }\bibfield  {title} {\bibinfo {title} {{Optimized Norm-Conserving
  Vanderbilt Pseudopotentials}},\ }\href@noop {} {\bibfield  {journal}
  {\bibinfo  {journal} {Phys. Rev. B}\ }\textbf {\bibinfo {volume} {88}},\
  \bibinfo {pages} {085117} (\bibinfo {year} {2013})}\BibitemShut {NoStop}%
\bibitem [{\citenamefont {Wu}\ \emph {et~al.}(2021)\citenamefont {Wu},
  \citenamefont {Smart}, \citenamefont {Xu},\ and\ \citenamefont
  {Ping}}]{Tyler2021-err}%
  \BibitemOpen
  \bibfield  {author} {\bibinfo {author} {\bibfnamefont {F.}~\bibnamefont
  {Wu}}, \bibinfo {author} {\bibfnamefont {T.~J.}\ \bibnamefont {Smart}},
  \bibinfo {author} {\bibfnamefont {J.}~\bibnamefont {Xu}},\ and\ \bibinfo
  {author} {\bibfnamefont {Y.}~\bibnamefont {Ping}},\ }\bibfield  {title}
  {\bibinfo {title} {Erratum: Carrier recombination mechanism at defects in
  wide band gap two-dimensional materials from first principles [phys. rev. b
  100, 081407(r) (2019)]},\ }\href
  {https://doi.org/10.1103/PhysRevB.104.079901} {\bibfield  {journal} {\bibinfo
   {journal} {Phys. Rev. B}\ }\textbf {\bibinfo {volume} {104}},\ \bibinfo
  {pages} {079901} (\bibinfo {year} {2021})}\BibitemShut {NoStop}%
\bibitem [{\citenamefont {Smart}\ \emph {et~al.}(2021)\citenamefont {Smart},
  \citenamefont {Li}, \citenamefont {Xu},\ and\ \citenamefont
  {Ping}}]{Smart2021wk}%
  \BibitemOpen
  \bibfield  {author} {\bibinfo {author} {\bibfnamefont {T.~J.}\ \bibnamefont
  {Smart}}, \bibinfo {author} {\bibfnamefont {K.}~\bibnamefont {Li}}, \bibinfo
  {author} {\bibfnamefont {J.}~\bibnamefont {Xu}},\ and\ \bibinfo {author}
  {\bibfnamefont {Y.}~\bibnamefont {Ping}},\ }\bibfield  {title} {\bibinfo
  {title} {{Intersystem Crossing and Exciton{\textendash}Defect Coupling of
  Spin Defects in Hexagonal Boron Nitride}},\ }\href@noop {} {\bibfield
  {journal} {\bibinfo  {journal} {npj Comput. Mater.}\ }\textbf {\bibinfo
  {volume} {7}},\ \bibinfo {pages} {59} (\bibinfo {year} {2021})}\BibitemShut
  {NoStop}%
\bibitem [{\citenamefont {Sundararaman}\ and\ \citenamefont
  {Ping}(2017)}]{PingJCP}%
  \BibitemOpen
  \bibfield  {author} {\bibinfo {author} {\bibfnamefont {R.}~\bibnamefont
  {Sundararaman}}\ and\ \bibinfo {author} {\bibfnamefont {Y.}~\bibnamefont
  {Ping}},\ }\bibfield  {title} {\bibinfo {title} {First-principles
  electrostatic potentials for reliable alignment at interfaces and defects},\
  }\href {https://doi.org/10.1063/1.4978238} {\bibfield  {journal} {\bibinfo
  {journal} {J. Chem. Phys.}\ }\textbf {\bibinfo {volume} {146}},\ \bibinfo
  {pages} {104109} (\bibinfo {year} {2017})}\BibitemShut {NoStop}%
\bibitem [{\citenamefont {Wu}\ \emph {et~al.}(2017)\citenamefont {Wu},
  \citenamefont {Galatas}, \citenamefont {Sundararaman}, \citenamefont
  {Rocca},\ and\ \citenamefont {Ping}}]{wu2017first}%
  \BibitemOpen
  \bibfield  {author} {\bibinfo {author} {\bibfnamefont {F.}~\bibnamefont
  {Wu}}, \bibinfo {author} {\bibfnamefont {A.}~\bibnamefont {Galatas}},
  \bibinfo {author} {\bibfnamefont {R.}~\bibnamefont {Sundararaman}}, \bibinfo
  {author} {\bibfnamefont {D.}~\bibnamefont {Rocca}},\ and\ \bibinfo {author}
  {\bibfnamefont {Y.}~\bibnamefont {Ping}},\ }\bibfield  {title} {\bibinfo
  {title} {First-principles engineering of charged defects for two-dimensional
  quantum technologies},\ }\href@noop {} {\bibfield  {journal} {\bibinfo
  {journal} {Phys. Rev. Mater.}\ }\textbf {\bibinfo {volume} {1}},\ \bibinfo
  {pages} {071001} (\bibinfo {year} {2017})}\BibitemShut {NoStop}%
\bibitem [{\citenamefont {Sundararaman}\ \emph {et~al.}(2017)\citenamefont
  {Sundararaman}, \citenamefont {Letchworth-Weaver}, \citenamefont {Schwarz},
  \citenamefont {Gunceler}, \citenamefont {Ozhabes},\ and\ \citenamefont
  {Arias}}]{JDFTx}%
  \BibitemOpen
  \bibfield  {author} {\bibinfo {author} {\bibfnamefont {R.}~\bibnamefont
  {Sundararaman}}, \bibinfo {author} {\bibfnamefont {K.}~\bibnamefont
  {Letchworth-Weaver}}, \bibinfo {author} {\bibfnamefont {K.~A.}\ \bibnamefont
  {Schwarz}}, \bibinfo {author} {\bibfnamefont {D.}~\bibnamefont {Gunceler}},
  \bibinfo {author} {\bibfnamefont {Y.}~\bibnamefont {Ozhabes}},\ and\ \bibinfo
  {author} {\bibfnamefont {T.}~\bibnamefont {Arias}},\ }\bibfield  {title}
  {\bibinfo {title} {{JDFTx:} software for joint density-functional theory},\
  }\href@noop {} {\bibfield  {journal} {\bibinfo  {journal} {SoftwareX}\
  }\textbf {\bibinfo {volume} {6}},\ \bibinfo {pages} {278 } (\bibinfo {year}
  {2017})}\BibitemShut {NoStop}%
\bibitem [{\citenamefont {Perdew}\ \emph {et~al.}(1996)\citenamefont {Perdew},
  \citenamefont {Burke},\ and\ \citenamefont {Ernzerhof}}]{PBE1997}%
  \BibitemOpen
  \bibfield  {author} {\bibinfo {author} {\bibfnamefont {J.~P.}\ \bibnamefont
  {Perdew}}, \bibinfo {author} {\bibfnamefont {K.}~\bibnamefont {Burke}},\ and\
  \bibinfo {author} {\bibfnamefont {M.}~\bibnamefont {Ernzerhof}},\ }\bibfield
  {title} {\bibinfo {title} {Generalized gradient approximation made simple},\
  }\href@noop {} {\bibfield  {journal} {\bibinfo  {journal} {Phys. Rev. Lett.}\
  }\textbf {\bibinfo {volume} {77}},\ \bibinfo {pages} {3865} (\bibinfo {year}
  {1996})}\BibitemShut {NoStop}%
\bibitem [{\citenamefont {Marini}\ \emph {et~al.}(2009)\citenamefont {Marini},
  \citenamefont {Hogan}, \citenamefont {Grüning},\ and\ \citenamefont
  {Varsano}}]{YAMBO}%
  \BibitemOpen
  \bibfield  {author} {\bibinfo {author} {\bibfnamefont {A.}~\bibnamefont
  {Marini}}, \bibinfo {author} {\bibfnamefont {C.}~\bibnamefont {Hogan}},
  \bibinfo {author} {\bibfnamefont {M.}~\bibnamefont {Grüning}},\ and\
  \bibinfo {author} {\bibfnamefont {D.}~\bibnamefont {Varsano}},\ }\bibfield
  {title} {\bibinfo {title} {Yambo: An ab initio tool for excited state
  calculations},\ }\href@noop {} {\bibfield  {journal} {\bibinfo  {journal}
  {Comput. Phys. Commun.}\ }\textbf {\bibinfo {volume} {180}},\ \bibinfo
  {pages} {1392 } (\bibinfo {year} {2009})}\BibitemShut {NoStop}%
\bibitem [{\citenamefont {Fuchs}\ \emph {et~al.}(2007)\citenamefont {Fuchs},
  \citenamefont {Furthm{\"u}ller}, \citenamefont {Bechstedt}, \citenamefont
  {Shishkin},\ and\ \citenamefont {Kresse}}]{fuchs2007quasiparticle}%
  \BibitemOpen
  \bibfield  {author} {\bibinfo {author} {\bibfnamefont {F.}~\bibnamefont
  {Fuchs}}, \bibinfo {author} {\bibfnamefont {J.}~\bibnamefont
  {Furthm{\"u}ller}}, \bibinfo {author} {\bibfnamefont {F.}~\bibnamefont
  {Bechstedt}}, \bibinfo {author} {\bibfnamefont {M.}~\bibnamefont
  {Shishkin}},\ and\ \bibinfo {author} {\bibfnamefont {G.}~\bibnamefont
  {Kresse}},\ }\bibfield  {title} {\bibinfo {title} {Quasiparticle band
  structure based on a generalized kohn-sham scheme},\ }\href@noop {}
  {\bibfield  {journal} {\bibinfo  {journal} {Phys. Rev. B}\ }\textbf {\bibinfo
  {volume} {76}},\ \bibinfo {pages} {115109} (\bibinfo {year}
  {2007})}\BibitemShut {NoStop}%
\bibitem [{\citenamefont {Bechstedt}(2016)}]{bechstedt2016many}%
  \BibitemOpen
  \bibfield  {author} {\bibinfo {author} {\bibfnamefont {F.}~\bibnamefont
  {Bechstedt}},\ }\href@noop {} {\emph {\bibinfo {title} {Many-Body Approach to
  Electronic Excitations}}}\ (\bibinfo  {publisher} {Springer-Verlag: Berlin},\
  \bibinfo {year} {2016})\BibitemShut {NoStop}%
\bibitem [{\citenamefont {Godby}\ and\ \citenamefont
  {Needs}(1989)}]{godby1989metal}%
  \BibitemOpen
  \bibfield  {author} {\bibinfo {author} {\bibfnamefont {R.~W.}\ \bibnamefont
  {Godby}}\ and\ \bibinfo {author} {\bibfnamefont {R.}~\bibnamefont {Needs}},\
  }\bibfield  {title} {\bibinfo {title} {Metal-insulator transition in
  kohn-sham theory and quasiparticle theory},\ }\href@noop {} {\bibfield
  {journal} {\bibinfo  {journal} {Phys. Rev. Lett.}\ }\textbf {\bibinfo
  {volume} {62}},\ \bibinfo {pages} {1169} (\bibinfo {year}
  {1989})}\BibitemShut {NoStop}%
\bibitem [{\citenamefont {Oschlies}\ \emph {et~al.}(1995)\citenamefont
  {Oschlies}, \citenamefont {Godby},\ and\ \citenamefont
  {Needs}}]{oschlies1995gw}%
  \BibitemOpen
  \bibfield  {author} {\bibinfo {author} {\bibfnamefont {A.}~\bibnamefont
  {Oschlies}}, \bibinfo {author} {\bibfnamefont {R.}~\bibnamefont {Godby}},\
  and\ \bibinfo {author} {\bibfnamefont {R.}~\bibnamefont {Needs}},\ }\bibfield
   {title} {\bibinfo {title} {Gw self-energy calculations of carrier-induced
  band-gap narrowing in n-type silicon},\ }\href@noop {} {\bibfield  {journal}
  {\bibinfo  {journal} {Phys. Rev. B}\ }\textbf {\bibinfo {volume} {51}},\
  \bibinfo {pages} {1527} (\bibinfo {year} {1995})}\BibitemShut {NoStop}%
\bibitem [{\citenamefont {Rozzi}\ \emph {et~al.}(2006)\citenamefont {Rozzi},
  \citenamefont {Varsano}, \citenamefont {Marini}, \citenamefont {Gross},\ and\
  \citenamefont {Rubio}}]{rozzi2006exact}%
  \BibitemOpen
  \bibfield  {author} {\bibinfo {author} {\bibfnamefont {C.~A.}\ \bibnamefont
  {Rozzi}}, \bibinfo {author} {\bibfnamefont {D.}~\bibnamefont {Varsano}},
  \bibinfo {author} {\bibfnamefont {A.}~\bibnamefont {Marini}}, \bibinfo
  {author} {\bibfnamefont {E.~K.}\ \bibnamefont {Gross}},\ and\ \bibinfo
  {author} {\bibfnamefont {A.}~\bibnamefont {Rubio}},\ }\bibfield  {title}
  {\bibinfo {title} {Exact coulomb cutoff technique for supercell
  calculations},\ }\href@noop {} {\bibfield  {journal} {\bibinfo  {journal}
  {Phys. Rev. B}\ }\textbf {\bibinfo {volume} {73}},\ \bibinfo {pages} {205119}
  (\bibinfo {year} {2006})}\BibitemShut {NoStop}%
\bibitem [{\citenamefont {Alkauskas}\ \emph
  {et~al.}(2014{\natexlab{a}})\citenamefont {Alkauskas}, \citenamefont {Yan},\
  and\ \citenamefont {Van~de Walle}}]{nonrad_alkauskas2014first}%
  \BibitemOpen
  \bibfield  {author} {\bibinfo {author} {\bibfnamefont {A.}~\bibnamefont
  {Alkauskas}}, \bibinfo {author} {\bibfnamefont {Q.}~\bibnamefont {Yan}},\
  and\ \bibinfo {author} {\bibfnamefont {C.~G.}\ \bibnamefont {Van~de Walle}},\
  }\bibfield  {title} {\bibinfo {title} {First-principles theory of
  nonradiative carrier capture via multiphonon emission},\ }\href@noop {}
  {\bibfield  {journal} {\bibinfo  {journal} {Phys. Rev. B}\ }\textbf {\bibinfo
  {volume} {90}},\ \bibinfo {pages} {075202} (\bibinfo {year}
  {2014}{\natexlab{a}})}\BibitemShut {NoStop}%
\bibitem [{\citenamefont {Alkauskas}\ \emph
  {et~al.}(2014{\natexlab{b}})\citenamefont {Alkauskas}, \citenamefont
  {Buckley}, \citenamefont {Awschalom},\ and\ \citenamefont {Van~de
  Walle}}]{pl_alkauskas2014first}%
  \BibitemOpen
  \bibfield  {author} {\bibinfo {author} {\bibfnamefont {A.}~\bibnamefont
  {Alkauskas}}, \bibinfo {author} {\bibfnamefont {B.~B.}\ \bibnamefont
  {Buckley}}, \bibinfo {author} {\bibfnamefont {D.~D.}\ \bibnamefont
  {Awschalom}},\ and\ \bibinfo {author} {\bibfnamefont {C.~G.}\ \bibnamefont
  {Van~de Walle}},\ }\bibfield  {title} {\bibinfo {title} {First-principles
  theory of the luminescence lineshape for the triplet transition in diamond nv
  centres},\ }\href@noop {} {\bibfield  {journal} {\bibinfo  {journal} {New J.
  Phys.}\ }\textbf {\bibinfo {volume} {16}},\ \bibinfo {pages} {073026}
  (\bibinfo {year} {2014}{\natexlab{b}})}\BibitemShut {NoStop}%
\bibitem [{\citenamefont {Palummo}\ \emph {et~al.}(2015)\citenamefont
  {Palummo}, \citenamefont {Bernardi},\ and\ \citenamefont
  {Grossman}}]{palummo2015exciton}%
  \BibitemOpen
  \bibfield  {author} {\bibinfo {author} {\bibfnamefont {M.}~\bibnamefont
  {Palummo}}, \bibinfo {author} {\bibfnamefont {M.}~\bibnamefont {Bernardi}},\
  and\ \bibinfo {author} {\bibfnamefont {J.~C.}\ \bibnamefont {Grossman}},\
  }\bibfield  {title} {\bibinfo {title} {Exciton radiative lifetimes in
  two-dimensional transition metal dichalcogenides},\ }\href@noop {} {\bibfield
   {journal} {\bibinfo  {journal} {Nano Lett.}\ }\textbf {\bibinfo {volume}
  {15}},\ \bibinfo {pages} {2794} (\bibinfo {year} {2015})}\BibitemShut
  {NoStop}%
\bibitem [{\citenamefont {Guo}\ \emph {et~al.}(2021)\citenamefont {Guo},
  \citenamefont {Xu},\ and\ \citenamefont {Ping}}]{guo2021substrate}%
  \BibitemOpen
  \bibfield  {author} {\bibinfo {author} {\bibfnamefont {C.}~\bibnamefont
  {Guo}}, \bibinfo {author} {\bibfnamefont {J.}~\bibnamefont {Xu}},\ and\
  \bibinfo {author} {\bibfnamefont {Y.}~\bibnamefont {Ping}},\ }\bibfield
  {title} {\bibinfo {title} {Substrate effect on excitonic shift and radiative
  lifetime of two-dimensional materials},\ }\href@noop {} {\bibfield  {journal}
  {\bibinfo  {journal} {J. of Phys. Condens. Matter}\ }\textbf {\bibinfo
  {volume} {33}},\ \bibinfo {pages} {234001} (\bibinfo {year}
  {2021})}\BibitemShut {NoStop}%
\bibitem [{\citenamefont {Attaccalite}\ \emph {et~al.}(2011)\citenamefont
  {Attaccalite}, \citenamefont {Bockstedte}, \citenamefont {Marini},
  \citenamefont {Rubio},\ and\ \citenamefont {Wirtz}}]{Attaccalite2011}%
  \BibitemOpen
  \bibfield  {author} {\bibinfo {author} {\bibfnamefont {C.}~\bibnamefont
  {Attaccalite}}, \bibinfo {author} {\bibfnamefont {M.}~\bibnamefont
  {Bockstedte}}, \bibinfo {author} {\bibfnamefont {A.}~\bibnamefont {Marini}},
  \bibinfo {author} {\bibfnamefont {A.}~\bibnamefont {Rubio}},\ and\ \bibinfo
  {author} {\bibfnamefont {L.}~\bibnamefont {Wirtz}},\ }\bibfield  {title}
  {\bibinfo {title} {Coupling of excitons and defect states in boron-nitride
  nanostructures},\ }\href {https://doi.org/10.1103/PhysRevB.83.144115}
  {\bibfield  {journal} {\bibinfo  {journal} {Phys. Rev. B}\ }\textbf {\bibinfo
  {volume} {83}},\ \bibinfo {pages} {144115} (\bibinfo {year}
  {2011})}\BibitemShut {NoStop}%
\bibitem [{\citenamefont {Smart}\ \emph {et~al.}(2018)\citenamefont {Smart},
  \citenamefont {Wu}, \citenamefont {Govoni},\ and\ \citenamefont
  {Ping}}]{smart2018}%
  \BibitemOpen
  \bibfield  {author} {\bibinfo {author} {\bibfnamefont {T.~J.}\ \bibnamefont
  {Smart}}, \bibinfo {author} {\bibfnamefont {F.}~\bibnamefont {Wu}}, \bibinfo
  {author} {\bibfnamefont {M.}~\bibnamefont {Govoni}},\ and\ \bibinfo {author}
  {\bibfnamefont {Y.}~\bibnamefont {Ping}},\ }\bibfield  {title} {\bibinfo
  {title} {Fundamental principles for calculating charged defect ionization
  energies in ultrathin two-dimensional materials},\ }\href@noop {} {\bibfield
  {journal} {\bibinfo  {journal} {Phys. Rev. Mater.}\ }\textbf {\bibinfo
  {volume} {2}},\ \bibinfo {pages} {124002} (\bibinfo {year}
  {2018})}\BibitemShut {NoStop}%
\bibitem [{\citenamefont {Novotny}\ and\ \citenamefont
  {Hecht}(2012)}]{novotny2012principles}%
  \BibitemOpen
  \bibfield  {author} {\bibinfo {author} {\bibfnamefont {L.}~\bibnamefont
  {Novotny}}\ and\ \bibinfo {author} {\bibfnamefont {B.}~\bibnamefont
  {Hecht}},\ }\href@noop {} {\emph {\bibinfo {title} {Principles of
  Nano-Optics}}}\ (\bibinfo  {publisher} {Cambridge university press},\
  \bibinfo {year} {2012})\BibitemShut {NoStop}%
\bibitem [{\citenamefont {Preu{\ss}}\ \emph {et~al.}(2021)\citenamefont
  {Preu{\ss}}, \citenamefont {Rudi}, \citenamefont {Kern}, \citenamefont
  {Schmidt}, \citenamefont {Bratschitsch},\ and\ \citenamefont
  {de~Vasconcellos}}]{preuss2021assembly}%
  \BibitemOpen
  \bibfield  {author} {\bibinfo {author} {\bibfnamefont {J.~A.}\ \bibnamefont
  {Preu{\ss}}}, \bibinfo {author} {\bibfnamefont {E.}~\bibnamefont {Rudi}},
  \bibinfo {author} {\bibfnamefont {J.}~\bibnamefont {Kern}}, \bibinfo {author}
  {\bibfnamefont {R.}~\bibnamefont {Schmidt}}, \bibinfo {author} {\bibfnamefont
  {R.}~\bibnamefont {Bratschitsch}},\ and\ \bibinfo {author} {\bibfnamefont
  {S.~M.}\ \bibnamefont {de~Vasconcellos}},\ }\bibfield  {title} {\bibinfo
  {title} {Assembly of large hbn nanocrystal arrays for quantum light
  emission},\ }\href@noop {} {\bibfield  {journal} {\bibinfo  {journal} {2D
  Mater.}\ }\textbf {\bibinfo {volume} {8}},\ \bibinfo {pages} {035005}
  (\bibinfo {year} {2021})}\BibitemShut {NoStop}%
\bibitem [{\citenamefont {Auburger}\ and\ \citenamefont
  {Gali}(2021)}]{auburger2021towards}%
  \BibitemOpen
  \bibfield  {author} {\bibinfo {author} {\bibfnamefont {P.}~\bibnamefont
  {Auburger}}\ and\ \bibinfo {author} {\bibfnamefont {A.}~\bibnamefont
  {Gali}},\ }\bibfield  {title} {\bibinfo {title} {Towards ab initio
  identification of paramagnetic substitutional carbon defects in hexagonal
  boron nitride acting as quantum bits},\ }\href@noop {} {\bibfield  {journal}
  {\bibinfo  {journal} {Phys. Rev. B}\ }\textbf {\bibinfo {volume} {104}},\
  \bibinfo {pages} {075410} (\bibinfo {year} {2021})}\BibitemShut {NoStop}%
\bibitem [{\citenamefont {Towns}\ \emph {et~al.}(2014)\citenamefont {Towns},
  \citenamefont {Cockerill}, \citenamefont {Dahan}, \citenamefont {Foster},
  \citenamefont {Gaither}, \citenamefont {Grimshaw}, \citenamefont {Hazlewood},
  \citenamefont {Lathrop}, \citenamefont {Lifka}, \citenamefont {Peterson},
  \citenamefont {Roskies}, \citenamefont {Scott},\ and\ \citenamefont
  {Wilkins-Diehr}}]{xsede}%
  \BibitemOpen
  \bibfield  {author} {\bibinfo {author} {\bibfnamefont {J.}~\bibnamefont
  {Towns}}, \bibinfo {author} {\bibfnamefont {T.}~\bibnamefont {Cockerill}},
  \bibinfo {author} {\bibfnamefont {M.}~\bibnamefont {Dahan}}, \bibinfo
  {author} {\bibfnamefont {I.}~\bibnamefont {Foster}}, \bibinfo {author}
  {\bibfnamefont {K.}~\bibnamefont {Gaither}}, \bibinfo {author} {\bibfnamefont
  {A.}~\bibnamefont {Grimshaw}}, \bibinfo {author} {\bibfnamefont
  {V.}~\bibnamefont {Hazlewood}}, \bibinfo {author} {\bibfnamefont
  {S.}~\bibnamefont {Lathrop}}, \bibinfo {author} {\bibfnamefont
  {D.}~\bibnamefont {Lifka}}, \bibinfo {author} {\bibfnamefont {G.~D.}\
  \bibnamefont {Peterson}}, \bibinfo {author} {\bibfnamefont {R.}~\bibnamefont
  {Roskies}}, \bibinfo {author} {\bibfnamefont {J.~R.}\ \bibnamefont {Scott}},\
  and\ \bibinfo {author} {\bibfnamefont {N.}~\bibnamefont {Wilkins-Diehr}},\
  }\bibfield  {title} {\bibinfo {title} {Xsede: Accelerating scientific
  discovery},\ }\href@noop {} {\bibfield  {journal} {\bibinfo  {journal}
  {Comput. Sci. Eng.}\ }\textbf {\bibinfo {volume} {16}},\ \bibinfo {pages}
  {62} (\bibinfo {year} {2014})}\BibitemShut {NoStop}%
\end{thebibliography}%
\end{document}